\documentstyle[11pt]{article}

\begin{document}

\author{Gediminas Gaigalas, Zenonas Rudzikas  \\ 
{\em State Institute of Theoretical Physics and Astronomy,} \\ {\em A. Go\v{s%
}tauto 12, 2600 Vilnius, LITHUANIA}}
\date{}
\title{On the secondly quantized theory of many-electron atom}
\maketitle

\begin{abstract}
Traditional theory of many-electron atoms and ions is based on the
coefficients of fractional parentage and matrix elements of tensorial
operators, composed of unit tensors. Then the calculation of spin-angular
coefficients of radial integrals appearing in the expressions of matrix
elements of arbitrary physical operators of atomic quantities has two main
disadvantages: (i) The numerical codes for the calculation of spin-angular
coefficients are usually very time-consuming; (ii) f-shells are often omitted
from programs for matrix element calculation since the tables for their
coefficients of fractional parentage are very extensive.

The authors suppose that a series of difficulties persisting in the
traditional approach to the calculation of spin-angular parts of matrix
elements could be avoided by using this secondly quantized methodology,
based on angular momentum theory, on the concept of the irreducible
tensorial sets, on a generalized graphical method, on quasispin and on the
reduced coefficients of fractional parentage.
\end{abstract}

\vspace{1.in} On the secondly quantized theory of many-electron atom

{\bf PACS: 3110, 3115}

\clearpage

\clearpage

\section{\bf Introduction}

Modern atomic spectroscopy studies the structure and properties of
practically any atom of the periodic table as well as of ions of any
ionization degree. Particular attention is paid to their energy spectra. For
the investigations of many-electron atoms and ions, it is of great
importance to combine experimental and theoretical methods. Nowadays the
possibilities of theoretical spectroscopy are much enlarged thanks to the
wide usage of powerful computers. Theoretical methods utilized must be
fairly universal and must ensure reasonably accurate values of physical
quantities studied.

Many-electron atom usually is considered as many-body problem and is
described by the wave function constructed from the wave functions of one
electron, moving in the central nuclear charge field and in the screening
field of the remaining electrons. Then the wave function of this electron
may be represented as a product of radial and spin-angular parts. The radial
part is usually found by solving various modifications of the Hartree-Fock
equations and can be represented in a numerical or analytical forms (Froese
Fischer 1977) whereas the angular part is expressed in terms of spherical
functions. Then the wave function of the whole atom can be constructed in
some standard way (Cowan 1981, Jucys and Savukynas 1973, Nikitin and
Rudzikas 1983) starting with these one-electron functions and may be used
further on for the calculations of any matrix elements representing physical
quantities.

During the last two decades a number of new versions of the technique
(so-called Racah algebra) to cope with spin-angular parts of the wave
functions and matrix elements have been suggested (Rudzikas 1991). Among
them the second quantization and quasispin techniques turned out to be of
particular efficiency (Judd 1967, Rudzikas and Kaniauskas 1984). The usage
of graphical methods (Jucys and Bandzaitis 1977) allowed one to find general
expressions even for rather complex cases of matrix elements. All this
enabled one to formulate fairly consistent and general non-relativistic and
relativistic theory of many-electron atom and processes of its interaction
with electromagnetic radiation (Rudzikas 1996). The abovementioned methods
are applicable for the both variational and perturbative approaches for
various coupling schemes of spin and orbital momenta.

Practically we have to solve so-called eigenvalue problem

\begin{equation}
\label{eq:Sch}H\Psi =E\Psi ,
\end{equation}
where $\Psi $ is the wave function of the system under investigation and $H$
is its Hamiltonian. In various versions of perturbation theory such equation
usually serves as the starting point for the further refinements. It turned
out that for very large variety of atoms and their ionization degrees the
so-called Hartree-Fock-Pauli Hamiltonian leads to highly accurate energy
values (Nikitin and Rudzikas 1983, Rudzikas 1996) that is why it is widely
used in many methods and computer codes.

In order to calculate the energy spectrum of atom or ion we have to find the
expressions for the matrix elements of all terms of the Hamiltonian
considered. For complex electronic configurations, having several open
shells, this is a task very far from the trivial one. For the optimization
of their expressions one has to combine the methods of the angular momentum
theory, irreducible tensorial sets, tensorial products in a coupled form,
coefficients of fractional parentage with the utilization of the graphical
(diagrammatic) methods, second quantization and with the accounting for the
symmetry properties of the system under consideration in the additional
spaces, for example, quasispin space. This paper describes one such
possibility.

Unfortunately, practical calculations show that all realistic atomic
Hamiltonians do not lead straightforwardly to eigenvalue problem (\ref
{eq:Sch}). Actually we have to calculate all non-zero matrix elements of the
Hamiltonian considered including those non-diagonal with respect to
electronic configurations, then to form energy matrix, to diagonalize it,
obtaining in this way the values of the energy levels as well as the
eigenfunctions (the wave functions in the intermediate coupling scheme). The
latter may be used then to calculate electronic transitions as well as the
other properties and processes. Such necessity raises special requirements
to the theory.

The total matrix element of each term of the energy operator in the case of
complex electronic configuration will consist of matrix elements, describing
the interaction inside each shell (in relativistic case - each subshell) of
equivalent electrons as well as between these shells. Going beyond the
single-configuration approximation we have to be able to take into account
in the same way non-diagonal, with respect to configurations, matrix
elements. Starting at the very beginning with the second quantization and
quasispin methods we are in a position to fulfill all these requirements.
Below we shall describe the approach suggested in the more details.

\section{\bf Tensorial Form of the Operators}

According to the method of second quantization (Judd 1967, Rudzikas and
Kaniauskas 1984) any one-particle operator

\begin{equation}
\label{eq:fa}F=\displaystyle
{\sum_{i,j}}a_ia_j^{+}\left( i|f|j\right) ,
\end{equation}
can be expressed in the following tensorial form:
\begin{equation}
\label{eq:fb}F=\displaystyle
{\sum_{n_il_i,n_jl_j}}F(i,j)=\displaystyle
{\sum_{n_il_i,n_jl_j}}\left[ \kappa ,\sigma \right] ^{-1/2}\left( n_i\lambda
_i||f^{\left( \kappa \sigma \right) }||n_j\lambda _j\right) \left[ a^{\left(
\lambda _i\right) }\times \stackrel{\sim }{a}^{\left( \lambda _j\right)
}\right] _{m_\Gamma }^{\left( \kappa \sigma \right) \Gamma },
\end{equation}
where $i\equiv n_il_ism_{l_i}m_{s_i},\lambda \equiv ls,\left[ \kappa ,\sigma
\right] \equiv \left( 2\kappa +1\right) \left( 2\sigma +1\right) ,$ $\left(
n_i\lambda _i||f^{\left( \kappa \sigma \right) }||n_j\lambda _j\right) $ is
the one-electron submatrix (reduced matrix) element of operator $F$, and $%
a^{\left( \lambda _i\right) }$ is the electron creation operator. The tensor
$\stackrel{\sim }{a}^{\left( \lambda _j\right) }$ is defined as

\begin{equation}
\label{eq:fc}\stackrel{\sim }{a}_{m_\lambda }^{\left( \lambda \right)
}=\left( -1\right) ^{\lambda -m_\lambda }a_{-m_\lambda }^{\left( \lambda
\right) +},
\end{equation}
where $a_{-m_\lambda }^{\left( \lambda \right) +}$ is the electron
annihilation operator. From tensorial point of view it is better to consider
tensor $\stackrel{\sim }{a}^{\left( \lambda _j\right) }$ as electron
annihilation operator (see Section 4). The product of tensors $\left[
a^{\left( \lambda _i\right) }\times \stackrel{\sim }{a}^{\left( \lambda
_j\right) }\right] _{m_\Gamma }^{\left( \kappa \sigma \right) \Gamma }$
denotes tensorial part of operator $F$. Here the rank $\kappa $ of orbital
space is coupled to the spin space rank $\sigma $ to form a tensorial
product of total spin-angular rank $\Gamma $. As we shall see, this
expression is very effective for the calculation of spin-angular
coefficients for any one-particle operator. This expression is a general one
and the tensorial form of any one-particle physical operator may be obtained
from it. For example, the spin-orbit interaction operator has the tensorial
structure $\kappa =1,\sigma =1,\Gamma =0$, and its submatrix element is

\begin{equation}
\label{eq:fd}\left( n_i\lambda _i||f_{s-o}^{\left( 11\right) }||n_j\lambda
_j\right) =-z\alpha ^2\left( \frac 38l_i\left( l_i+1\right) \left(
2l_i+1\right) \right) ^{1/2}\left( n_il_i|1/r^3|n_jl_j\right) \delta \left(
l_i,l_j\right) .
\end{equation}

Any two-particle tensorial operator

\begin{equation}
\label{eq:ga}G=\frac 12\displaystyle
{\sum_{i,j,i^{\prime },j^{\prime }}}a_ia_ja_{j^{\prime }}^{+}a_{i^{\prime
}}^{+}\left( i,j|g|i^{\prime },j^{\prime }\right)
\end{equation}
can\- be expressed in two well-known forms (Rudzikas and Kaniauskas 1984).
In the first form the operators of second quantization follow in the normal
order:

\begin{equation}
\label{eq:gb}
\begin{array}[b]{c}
G_I=
\displaystyle {\sum_{n_il_i,n_jl_j,n_i^{\prime }l_i^{\prime },n_j^{\prime
}l_j^{\prime }}}G_I(iji^{\prime }j^{\prime })= \\ =-\frac 12
\displaystyle {\sum_{n_il_i,n_jl_j,n_i^{\prime }l_i^{\prime },n_j^{\prime
}l_j^{\prime }}}\displaystyle {\sum_{\kappa _{12}\kappa _{12}^{^{\prime
}}\sigma _{12}\sigma _{12}^{^{\prime }}}}\displaystyle {\sum_p}\left(
-1\right) ^{k-p}\left[ \kappa _{12},\kappa _{12}^{\prime },\sigma
_{12},\sigma _{12}^{\prime }\right] ^{1/2}\times \\ \times \left( n_i\lambda
_in_j\lambda _j||g^{\left( \kappa _1\kappa _2k,\sigma _1\sigma _2k\right)
}||n_i^{\prime }\lambda _i^{\prime }n_j^{\prime }\lambda _j^{\prime }\right)
\times \\
\times \left\{
\begin{array}{ccc}
l_i^{\prime } & l_j^{\prime } & \kappa _{12}^{\prime } \\
\kappa _1 & \kappa _2 & k \\
l_i & l_j & \kappa _{12}
\end{array}
\right\} \left\{
\begin{array}{ccc}
s & s & \sigma _{12}^{\prime } \\
\sigma _1 & \sigma _2 & k \\
s & s & \sigma _{12}
\end{array}
\right\} \times \\
\times \left[ \left[ a^{\left( \lambda _i\right) }\times a^{\left( \lambda
_j\right) }\right] ^{\left( \kappa _{12}\sigma _{12}\right) }\times \left[
\stackrel{\sim }{a}^{\left( \lambda _i^{\prime }\right) }\times \stackrel{%
\sim }{a}^{\left( \lambda _j^{\prime }\right) }\right] ^{\left( \kappa
_{12}^{\prime }\sigma _{12}^{\prime }\right) }\right] _{p-p}^{\left(
kk\right) },
\end{array}
\end{equation}
where $\left( n_i\lambda _in_j\lambda _j||g^{\left( \kappa _1\kappa
_2k,\sigma _1\sigma _2k\right) }||n_i^{\prime }\lambda _i^{\prime
}n_j^{\prime }\lambda _j^{\prime }\right) $ is the two-electron submatrix
element of operator $G$.

In another form the second quantization operators are coupled by pairs
consisting of electron creation and annihilation operators. In tensorial
form:

\begin{equation}
\label{eq:gc}
\begin{array}[b]{c}
G_{II}=
\displaystyle {\sum_{n_il_i,n_jl_j,n_i^{\prime }l_i^{\prime },n_j^{\prime
}l_j^{\prime }}}G_{II}(iji^{\prime }j^{\prime })= \\ =\frac 12
\displaystyle {\sum_{n_il_i,n_jl_j,n_i^{\prime }l_i^{\prime },n_j^{\prime
}l_j^{\prime }}}\displaystyle {\sum_p}\left( -1\right) ^{k-p}\left(
n_i\lambda _in_j\lambda _j||g^{\left( \kappa _1\kappa _2k,\sigma _1\sigma
_2k\right) }||n_i^{\prime }\lambda _i^{\prime }n_j^{\prime }\lambda
_j^{\prime }\right) \times \\ \times \{\left[ \kappa _1,\kappa _2,\sigma
_1,\sigma _2\right] ^{-1/2}\times \left[ \left[ a^{\left( \lambda _i\right)
}\times
\stackrel{\sim }{a}^{\left( \lambda _i^{\prime }\right) }\right] ^{\left(
\kappa _1\sigma _1\right) }\times \left[ a^{\left( \lambda _j\right) }\times
\stackrel{\sim }{a}^{\left( \lambda _j^{\prime }\right) }\right] ^{\left(
\kappa _2\sigma _2\right) }\right] _{p-p}^{\left( kk\right) }- \\ -\left(
-1\right) ^{l_i+l_j^{\prime }}\left\{
\begin{array}{ccc}
\kappa _1 & \kappa _2 & k \\
l_j^{\prime } & l_i & l_j
\end{array}
\right\} \left\{
\begin{array}{ccc}
\sigma _1 & \sigma _2 & k \\
s & s & s
\end{array}
\right\} \left[ a^{\left( \lambda _i\right) }\times \stackrel{\sim }{a}%
^{\left( \lambda _j^{\prime }\right) }\right] _{p-p}^{\left( kk\right)
}\delta \left( n_jl_j,n_i^{\prime }l_i^{\prime }\right) \}.
\end{array}
\end{equation}

The expression (\ref{eq:gb}) consists of only one tensorial product whereas (%
\ref{eq:gc}) has two, but the summation in the first formula is also over
intermediate ranks $\kappa _{12}$, $\sigma _{12}$, $\kappa _{12}^{\prime }$
and $\sigma _{12}^{\prime }$, complicating in this way the calculations. The
advantages or disadvantages of these alternative forms of arbitrary
two-electron operator may be revealed in practical applications.

In these forms the product of second quantization operators denotes
tensorial part of operator G. For instance, the tensorial structure of
electrostatic (Coulomb) electron interaction operator is the same as that of
orbit-orbit interaction, $\kappa _1=\kappa _2=k,\sigma _1=\sigma _2=0$
(Jucys and Savukynas 1984), and only the two-electron submatrix elements $%
\left( n_i\lambda _in_j\lambda _j||g^{\left( \kappa _1\kappa _2k,\sigma
_1\sigma _2k\right) }||n_i^{\prime }\lambda _i^{\prime }n_j^{\prime }\lambda
_j^{\prime }\right) $ of these operators are different. In the case of
electrostatic interaction:

\begin{equation}
\label{eq:gd}
\begin{array}[b]{c}
\left( n_i\lambda _in_j\lambda _j||g_{Coulomb}^{\left( kk0,000\right)
}||n_i^{\prime }\lambda _i^{\prime }n_j^{\prime }\lambda _j^{\prime }\right)
= \\
=2\left[ k\right] ^{1/2}\left( l_i||C^{\left( k\right) }||l_i^{\prime
}\right) \left( l_j||C^{\left( k\right) }||l_j^{\prime }\right) R_k\left(
n_il_in_i^{\prime }l_i^{\prime },n_jl_jn_j^{\prime }l_j^{\prime }\right) .
\end{array}
\end{equation}

From (\ref{eq:gd}), by (\ref{eq:gb}) and (\ref{eq:gc}), we finally obtain
the following two secondly quantized expressions for Coulomb operator:

\begin{equation}
\label{eq:ggaa}
\begin{array}[b]{c}
V_I=-\frac 12
\displaystyle {\sum_{n_il_in_jl_jn_i^{\prime }l_i^{\prime }n_j^{\prime
}l_j^{\prime }}}\displaystyle {\sum_{\kappa _{12}\sigma _{12}k}}\left(
-1\right) ^{l_j+l_i^{\prime }+k+\kappa _{12}}\left[ \kappa _{12},\sigma
_{12}\right] ^{1/2}\left( l_i||C^{\left( k\right) }||l_i^{\prime }\right)
\times \\ \times \left( l_j||C^{\left( k\right) }||l_j^{\prime }\right)
R_k\left( n_il_in_i^{\prime }l_i^{\prime },n_jl_jn_j^{\prime }l_j^{\prime
}\right) \left\{
\begin{array}{ccc}
l_i & l_i^{\prime } & k \\
l_j^{\prime } & l_i & \kappa _{12}
\end{array}
\right\} \times \\
\times \left[ \left[ a^{\left( \lambda _i\right) }\times a^{\left( \lambda
_j\right) }\right] ^{\left( \kappa _{12}\sigma _{12}\right) }\times \left[
\stackrel{\sim }{a}^{\left( \lambda _i^{\prime }\right) }\times \stackrel{%
\sim }{a}^{\left( \lambda _j^{\prime }\right) }\right] ^{\left( \kappa
_{12}\sigma _{12}\right) }\right] ^{\left( 00\right) },
\end{array}
\end{equation}
\begin{equation}
\label{eq:ggbbb}
\begin{array}[b]{c}
V_{II}=
\displaystyle {\sum_{n_il_in_jl_jn_i^{\prime }l_i^{\prime }n_j^{\prime
}l_j^{\prime }}}\displaystyle {\sum_k}\left( l_i||C^{\left( k\right)
}||l_i^{\prime }\right) \left( l_j||C^{\left( k\right) }||l_j^{\prime
}\right) R_k\left( n_il_in_i^{\prime }l_i^{\prime },n_jl_jn_j^{\prime
}l_j^{\prime }\right) \times \\ \times \{\left[ k\right] ^{-1/2}\left[
\left[ a^{\left( \lambda _i\right) }\times
\stackrel{\sim }{a}^{\left( \lambda _i^{\prime }\right) }\right] ^{\left(
k0\right) }\times \left[ a^{\left( \lambda _j\right) }\times \stackrel{\sim
}{a}^{\left( \lambda _j^{\prime }\right) }\right] ^{\left( k0\right)
}\right] ^{\left( 00\right) }+ \\ +\left( 2\left[ l_i\right] \right)
^{-1/2}\left[ a^{\left( \lambda _i\right) }\times \stackrel{\sim }{a}%
^{\left( \lambda _j^{\prime }\right) }\right] ^{\left( 00\right) }\delta
\left( n_jl_j,n_i^{\prime }l_i^{\prime }\right) \},
\end{array}
\end{equation}

The tensorial expressions for orbit-orbit and other physical operators in
second quantization form may be obtained in the same manner.

It is worth mentioning that the expressions (\ref{eq:ggaa}) and (\ref
{eq:ggbbb}) embrace, already in an operator form, the interaction terms both
the diagonal ones, relative to configurations, and the non-diagonal ones.
Non-diagonal terms define the interaction between all the possible electron
distributions over the configurations considered, differing by quantum
numbers not more than two electrons.

The merits of representing operators in one form or another (\ref{eq:ggaa})
or (\ref{eq:ggbbb}) are mostly determined by the technique used to find
their matrix elements and quantities in terms of which they are expressed.

\section{\bf Generalized Graphical Method}

In this paragraph we shall sketch the generalized version of graphical
technique, in which not only one- and two-particle operators are presented
in tensorial form (such graphs are analogical to Feynman-Goldstone diagrams
but they do not depend on magnetic quantum numbers (Merkelis {\it et al}
1986a, b), but which allows also to represent graphically any tensorial
product of the second quantization operators and to perform graphically the
operations with the secondly quantized operators as well as with their
tensorial products (Gaigalas {\it et al} 1985, Gaigalas 1985, Gaigalas and
Merkelis 1987). Such graphical technique is most suitable to represent any
one- and two-particle operator already presented in tensorial form and to
found general expressions for their matrix elements.

%
%
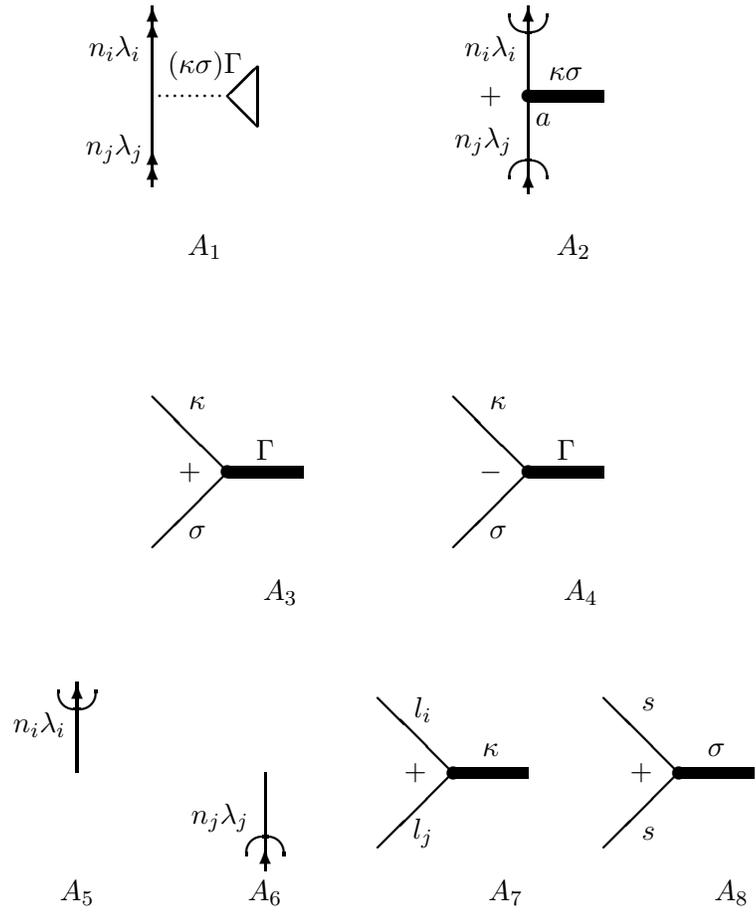
\begin{figure}
\setlength{\unitlength}{1mm}
\begin{picture}(148,135)
\thicklines
%
%
\put(20,130){\vector(0,1){10}}
\put(20,140){\vector(0,1){2}}
\put(15,138){\makebox(0,0)[t]{$ n_{i}\lambda_{i}$}}
\multiput(20,130)(1,0){10}{\circle*{0.02}}
\put(20,120){\line(0,1){10}}
\put(20,120){\vector(0,1){3}}
\put(15,123){\makebox(0,0){$ n_{j}\lambda_{j}$}}
\put(27,134){\makebox(0,0){$ (\kappa \sigma)\Gamma $}}
\put(20,118){\vector(0,1){3}}
\put(30,130){\line(1,1){4}}
\put(30,130){\line(1,-1){4}}
\put(34,126){\line(0,1){8}}
\put(27,110){\makebox(0,0){$ A_{1}$}}
%
%
\put(70,130){\vector(0,1){12}}
\put(70,141){\oval(5,5)[b]}
\put(70,130){\circle*{1,7}}
\put(65,138){\makebox(0,0)[t]{$ n_{i}\lambda_{i}$}}
\put(70,130.6){\line(10,0){10}}
\put(70,130.3){\line(10,0){10}}
\put(70,130){\line(10,0){10}}
\put(70,129.7){\line(10,0){10}}
\put(70,129.4){\line(10,0){10}}
\put(75,133){\makebox(0,0){$\kappa \sigma$}}
\put(70,120){\line(0,1){10}}
\put(64,124){\makebox(0,0){$ n_{j}\lambda_{j}$}}
\put(65,130){\makebox(0,0){${+}$}}
\put(72,127){\makebox(0,0){${a}$}}
\put(70,117){\vector(0,1){3}}
\put(70,119){\oval(5,5)[t]}
\put(76,110){\makebox(0,0){$ A_{2}$}}
%
%
\put(30,80){\line(-1,1){10}}
\put(30,80){\circle*{1,7}}
\put(26,90){\makebox(0,0)[t]{$ \kappa$}}
\put(30,80.6){\line(10,0){10}}
\put(30,80.3){\line(10,0){10}}
\put(30,80){\line(10,0){10}}
\put(30,79.7){\line(10,0){10}}
\put(30,79.4){\line(10,0){10}}
\put(35,83){\makebox(0,0){$\Gamma $}}
\put(30,80){\line(-1,-1){10}}
\put(26,72){\makebox(0,0){$ \sigma$}}
\put(25,80){\makebox(0,0){${+}$}}
\put(37,64){\makebox(0,0){$ A_{3}$}}
%
%
\put(70,80){\line(-1,1){10}}
\put(70,80){\circle*{1,7}}
\put(66,90){\makebox(0,0)[t]{$ \kappa$}}
\put(70,80.6){\line(10,0){10}}
\put(70,80.3){\line(10,0){10}}
\put(70,80){\line(10,0){10}}
\put(70,79.7){\line(10,0){10}}
\put(70,79.4){\line(10,0){10}}
\put(75,83){\makebox(0,0){$\Gamma $}}
\put(70,80){\line(-1,-1){10}}
\put(66,72){\makebox(0,0){$ \sigma$}}
\put(65,80){\makebox(0,0){${-}$}}
\put(77,64){\makebox(0,0){$ A_{4}$}}
%
%
\put(10,40){\vector(0,1){12}}
\put(10,51){\oval(5,5)[b]}
\put(5,48){\makebox(0,0)[t]{$ n_{i}\lambda_{i}$}}
\put(10,24){\makebox(0,0){$ A_{5}$}}
%
%
\put(35,30){\line(0,1){10}}
\put(29,34){\makebox(0,0){$ n_{j}\lambda_{j}$}}
\put(35,27){\vector(0,1){3}}
\put(35,29){\oval(5,5)[t]}
\put(35,24){\makebox(0,0){$ A_{6}$}}
%
%
\put(60,40){\line(-1,1){10}}
\put(60,40){\circle*{1,7}}
\put(56,50){\makebox(0,0)[t]{$ l_{i}$}}
\put(60,40.6){\line(10,0){10}}
\put(60,40.3){\line(10,0){10}}
\put(60,40){\line(10,0){10}}
\put(60,39.7){\line(10,0){10}}
\put(60,39.4){\line(10,0){10}}
\put(65,43){\makebox(0,0){$\kappa $}}
\put(60,40){\line(-1,-1){10}}
\put(56,32){\makebox(0,0){$ l_{j}$}}
\put(55,40){\makebox(0,0){${+}$}}
\put(67,24){\makebox(0,0){$ A_{7}$}}
%
%
\put(90,40){\line(-1,1){10}}
\put(90,40){\circle*{1,7}}
\put(86,50){\makebox(0,0)[t]{$ s$}}
\put(90,40.6){\line(10,0){10}}
\put(90,40.3){\line(10,0){10}}
\put(90,40){\line(10,0){10}}
\put(90,39.7){\line(10,0){10}}
\put(90,39.4){\line(10,0){10}}
\put(95,43){\makebox(0,0){$\sigma $}}
\put(90,40){\line(-1,-1){10}}
\put(86,32){\makebox(0,0){$ s$}}
\put(85,40){\makebox(0,0){${+}$}}
\put(97,24){\makebox(0,0){$ A_{8}$}}
\end{picture}
\caption{Diagrams for one-particle operators.}
\label{a-a}
\end{figure}
%
%

In this methodology the item under summation sign of the one-particle
operator (\ref{eq:fb}) has the following graphical form:

\begin{equation}
\label{eq;gr-a}F\left( i,j\right) =A_1=\left[ \kappa ,\sigma \right]
^{-1/2}\left( n_i\lambda _i||f^{\left( \kappa \sigma \right) }||n_j\lambda
_j\right) A_2A_3,
\end{equation}
where the diagrams $A_1$, $A_2$ and $A_3$ are presented on Figure 1. As we
see, the diagram of the operator itself, namely $A_1$, is similar to the
usual Feynman-Goldstone diagram (Lindgren and Morrison 1982), although here
the summation over magnetic quantum numbers $m_\lambda $ is performed. The
product of the diagrams $A_2$, $A_3$ represents the tensorial structure of
the operator:

\begin{equation}
\label{eq:gr-b}A_2A_3=\left[ a^{\left( \lambda _i\right) }\times \stackrel{%
\sim }{a}^{\left( \lambda _j\right) }\right] _{m_\Gamma }^{\left( \kappa
\sigma \right) \Gamma }=\displaystyle
{\sum_{m_\kappa ,m_\sigma }}\left[ a^{\left( \lambda _i\right) }\times
\stackrel{\sim }{a}^{\left( \lambda _j\right) }\right] _{m_\kappa m_\sigma
}^{\left( \kappa \sigma \right) }\left[
\begin{array}{ccc}
\kappa & \sigma & \Gamma \\
m_\kappa & m_\sigma & m_\Gamma
\end{array}
\right] ,
\end{equation}
where $A_2$ equals:

\begin{equation}
\label{eq:gr-c}A_2=\left[ a^{\left( \lambda _i\right) }\times \stackrel{\sim
}{a}^{\left( \lambda _j\right) }\right] _{m_\kappa m_\sigma }^{\left( \kappa
\sigma \right) },
\end{equation}
whereas $A_3$, by (Jucys and Bandzaitis 1977), is equal to:

\begin{equation}
\label{eq:gr-d}A_3=\left[
\begin{array}{ccc}
\kappa & \sigma & \Gamma \\
m_\kappa & m_\sigma & m_\Gamma
\end{array}
\right] .
\end{equation}

The heavy line in the diagram $A_3$ represents the resultant momentum $%
\Gamma $ whereas the plus sign of the vertex means that the momenta $\kappa $
and $\sigma $ are coupled into the resultant $\Gamma $ in counter-clockwise
direction. From the symmetry properties of the Clebsch-Gordan coefficients
the equality follows:

\begin{equation}
\label{eq:gr-da}A_3=\left( -1\right) ^{\kappa +\sigma -\Gamma }A_4.
\end{equation}

Then we can conclude that, if we change the sign of any vertex, then the
phase multiplier of the form $\left( -1\right) ^{\kappa +\sigma -\Gamma }$
occurs.

The electron creation operator $a^{\left( \lambda _i\right) }$ has the
following graphical form (Figure 1, $A_5$):

\begin{equation}
\label{eq:gr-e}a^{\left( \lambda _i\right) }=A_5,
\end{equation}
whereas $\stackrel{\sim }{a}^{\left( \lambda _j\right) }$

\begin{equation}
\label{eq:gr-f}\stackrel{\sim }{a}^{\left( \lambda _j\right) }=A_6.
\end{equation}

Thus, it is obvious that the diagram $A_2$ consists of the second
quantization operators $a^{\left( \lambda _i\right) }$ and $\stackrel{\sim }{%
a}^{\left( \lambda _j\right) }$ as well as of the Clebsch-Gordan coefficients

\begin{equation}
\label{eq:gr-g}A_7=\left[
\begin{array}{ccc}
l_i & l_j & \kappa \\
m_{l_i} & m_{l_j} & m_\kappa
\end{array}
\right] ,\;\;\;\;\;\;\;\;\;\;A_8=\left[
\begin{array}{ccc}
s & s & \sigma \\
m_s & m_s & m_\sigma
\end{array}
\right] ,
\end{equation}
which couple these operators into tensorial product and which may be
obtained from the diagram $A_2$ if to omit in them the graphical symbols of
the second quantization operators. It is necessary to bear in mind that,
while writing down the algebraic expression from the diagram $A_2$, always
in tensorial product there must be in the first place the second
quantization operator, which is above the vertex ''$a$'', whereas the second
place must occupy the operator, which is below the vertex ''$a$'' in the
diagram $A_2$. The scheme of their coupling into tensorial product is
defined by the sign of the vertex.

%
%
%
\begin{figure}
\setlength{\unitlength}{1mm}
\begin{picture}(148,105)
\thicklines
%
%
\put(10,90){\vector(0,1){10}}
\put(10,100){\vector(0,1){2}}
\put(5,98){\makebox(0,0)[t]{$ n_{i}\lambda_{i}$}}
\put(20,90){\vector(0,1){10}}
\put(20,100){\vector(0,1){2}}
\put(25,98){\makebox(0,0)[t]{$ n_{j}\lambda_{j}$}}
\multiput(11,90)(2,0){5}{\oval(2,1)[t]}
\put(15,93){\makebox(0,0){$kk$}}
\put(10,80){\line(0,1){10}}
\put(10,80){\vector(0,1){3}}
\put(5,83){\makebox(0,0){$ n'_{i}\lambda'_{i}$}}
\put(10,78){\vector(0,1){3}}
\put(20,80){\line(0,1){10}}
\put(20,80){\vector(0,1){3}}
\put(25,83){\makebox(0,0){$ n'_{j}\lambda'_{j}$}}
\put(20,78){\vector(0,1){3}}
\put(15,70){\makebox(0,0){$ B_{1}$}}
%
%
%
\put(45,90){\vector(0,1){10}}
\put(45,100){\vector(0,1){2}}
\put(40,98){\makebox(0,0)[t]{$ n_{i}\lambda_{i}$}}
\put(55,90){\vector(0,1){10}}
\put(55,100){\vector(0,1){2}}
\put(60,98){\makebox(0,0)[t]{$ n_{j}\lambda_{j}$}}
\multiput(45,90)(1,0){10}{\circle*{0.02}}
\put(50,93){\makebox(0,0){$kk$}}
\put(45,80){\line(0,1){10}}
\put(45,80){\vector(0,1){3}}
\put(40,83){\makebox(0,0){$ n'_{i}\lambda'_{i}$}}
\put(45,78){\vector(0,1){3}}
\put(55,80){\line(0,1){10}}
\put(55,80){\vector(0,1){3}}
\put(60,83){\makebox(0,0){$ n'_{j}\lambda'_{j}$}}
\put(55,78){\vector(0,1){3}}
\put(50,70){\makebox(0,0){$ B_{2}$}}
%
%
\put(80,90){\vector(0,1){12}}
\put(80,101){\oval(5,5)[b]}
\put(80,90){\circle*{1,7}}
\put(75,98){\makebox(0,0)[t]{$ n_{i}\lambda_{i}$}}
\put(110,90){\line(0,12){12}}
\put(110,102){\vector(0,-1){3}}
\put(110,101){\oval(5,5)[b]}
\put(110,90){\circle*{1,7}}
\put(115,98){\makebox(0,0)[t]{$ n'_{i}\lambda'_{i}$}}
\put(80,90.6){\line(6,0){6}}
\put(80,90.3){\line(6,0){6}}
\put(80,90){\line(20,0){30}}
\put(80,89.7){\line(6,0){6}}
\put(80,89.4){\line(6,0){6}}
\put(87,93){\makebox(0,0){$\kappa_{12} \sigma_{12}$}}
\put(104,90.6){\line(6,0){6}}
\put(104,90.3){\line(6,0){6}}
\put(104,89.7){\line(6,0){6}}
\put(104,89.4){\line(6,0){6}}
\put(103,93){\makebox(0,0){$\kappa_{12}' \sigma_{12}' $}}
\put(95,90){\circle*{1,7}}
\put(95,95){\makebox(0,0){${-}$}}
\put(95.6,85){\line(0,1){5}}
\put(95.3,85){\line(0,1){5}}
\put(95,85){\line(0,1){5}}
\put(94.7,85){\line(0,1){5}}
\put(94.4,85){\line(0,1){5}}
\put(95,82){\makebox(0,0){$k k$}}
\put(80,80){\line(0,1){10}}
\put(74,84){\makebox(0,0){$ n_{j}\lambda_{j}$}}
\put(75,90){\makebox(0,0){${+}$}}
\put(80,80){\vector(0,-1){3}}
\put(80,79){\oval(5,5)[t]}
\put(110,80){\line(0,1){10}}
\put(116,84){\makebox(0,0){$ n'_{j}\lambda'_{j}$}}
\put(115,90){\makebox(0,0){${-}$}}
\put(110,77){\vector(0,1){3}}
\put(110,79){\oval(5,5)[t]}
\put(95,70){\makebox(0,0){$ B_{3}$}}
%
%
\put(10,40){\vector(0,1){10}}
\put(10,50){\vector(0,1){2}}
\put(5,48){\makebox(0,0)[t]{$ n_{i}\lambda_{i}$}}
\multiput(10,40)(1,0){10}{\circle*{0.02}}
\put(15,37){\makebox(0,0){$kk$}}
\put(16,48){\makebox(0,0){$ i'=j$}}
\put(16,45){\vector(-1,0){2}}
\put(20,30){\line(0,1){10}}
\put(15,40){\oval(10,10)[t]}
\put(20,30){\vector(0,1){3}}
\put(26,33){\makebox(0,0){$ n'_{j}\lambda'_{j}$}}
\put(20,28){\vector(0,1){3}}
\put(15,20){\makebox(0,0){$ B_{4}$}}
%
%
\put(45,40){\vector(0,1){12}}
\put(45,51){\oval(5,5)[b]}
\put(45,40){\circle*{1,7}}
\put(40,48){\makebox(0,0)[t]{$ n_{i}\lambda_{i}$}}
\put(75,40){\vector(0,1){12}}
\put(75,51){\oval(5,5)[b]}
\put(75,40){\circle*{1,7}}
\put(80,48){\makebox(0,0)[t]{$ n_{j}\lambda_{j}$}}
\put(45,40.6){\line(6,0){6}}
\put(45,40.3){\line(6,0){6}}
\put(45,40){\line(20,0){30}}
\put(45,39.7){\line(6,0){6}}
\put(45,39.4){\line(6,0){6}}
\put(52,43){\makebox(0,0){$\kappa_{1} \sigma_{1}$}}
\put(69,40.6){\line(6,0){6}}
\put(69,40.3){\line(6,0){6}}
\put(69,39.7){\line(6,0){6}}
\put(69,39.4){\line(6,0){6}}
\put(68,43){\makebox(0,0){$\kappa_{2} \sigma_{2}$}}
\put(60,40){\circle*{1,7}}
\put(60,45){\makebox(0,0){${-}$}}
\put(60.6,35){\line(0,1){5}}
\put(60.3,35){\line(0,1){5}}
\put(60,35){\line(0,1){5}}
\put(59.7,35){\line(0,1){5}}
\put(59.4,35){\line(0,1){5}}
\put(60,32){\makebox(0,0){$k k$}}
\put(45,30){\line(0,1){10}}
\put(39,34){\makebox(0,0){$ n'_{i}\lambda'_{i}$}}
\put(40,40){\makebox(0,0){${+}$}}
\put(45,27){\vector(0,1){3}}
\put(45,29){\oval(5,5)[t]}
\put(75,30){\line(0,1){10}}
\put(81,34){\makebox(0,0){$ n'_{j}\lambda'_{j}$}}
\put(80,40){\makebox(0,0){${-}$}}
\put(75,27){\vector(0,1){3}}
\put(75,29){\oval(5,5)[t]}
\put(60,20){\makebox(0,0){$ B_{5}$}}
%
%
\put(110,42){\circle*{1,7}}
\put(110,37){\makebox(0,0){${+}$}}
\put(110.6,42){\line(0,1){5}}
\put(110.3,42){\line(0,1){5}}
\put(110,42){\line(0,1){5}}
\put(109.7,42){\line(0,1){5}}
\put(109.4,42){\line(0,1){5}}
\put(115,45){\makebox(0,0){$k k$}}
\put(100,32){\line(1,1){10}}
\put(100,27){\line(0,1){5}}
\put(94,34){\makebox(0,0){$ n_{i}\lambda_{i}$}}
\put(100,30){\vector(0,-1){3}}
\put(100,29){\oval(5,5)[t]}
\put(120,32){\line(-1,1){10}}
\put(120,27){\line(0,1){5}}
\put(126,34){\makebox(0,0){$ n'_{j}\lambda'_{j}$}}
\put(120,27){\vector(0,1){3}}
\put(120,29){\oval(5,5)[t]}
\put(110,20){\makebox(0,0){$ B_{6}$}}
\end{picture}
\caption{Diagrams for two-particle operators.}
\label{bb}
\end{figure}
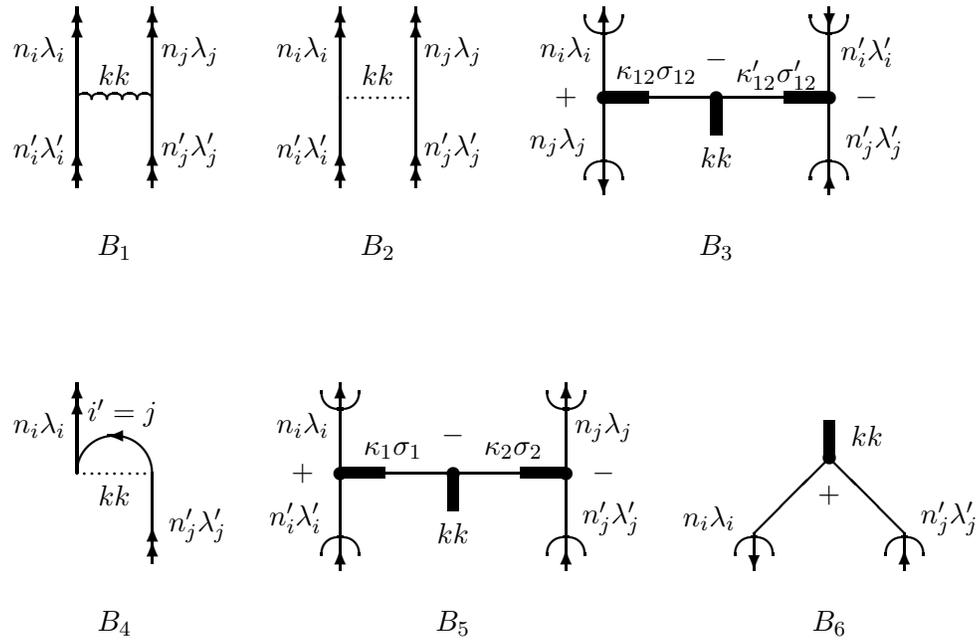
%
%

The first form (\ref{eq:gb}) of two-particle operator $G_I(iji^{\prime
}j^{\prime })$ is represented by the following diagram (Figure 2, $B_1$):

\begin{equation}
\label{eq:gr-h}
\begin{array}[b]{c}
G_I(iji^{\prime }j^{\prime })=B_1=-\frac 12
\displaystyle {\sum_{\kappa _{12}\kappa _{12}^{^{\prime }}\sigma _{12}\sigma
_{12}^{^{\prime }}}}\displaystyle {\sum_p}\left( -1\right) ^{k-p}\left[
\kappa _1,\kappa _2,\sigma _1,\sigma _2\right] ^{1/2}\times \\ \times \left(
n_i\lambda _in_j\lambda _j||g^{\left( \kappa _1\kappa _2k,\sigma _1\sigma
_2k\right) }||n_i^{\prime }\lambda _i^{\prime }n_j^{\prime }\lambda
_j^{\prime }\right) \left\{
\begin{array}{ccc}
l_i^{\prime } & l_j^{\prime } & \kappa _{12}^{\prime } \\
\kappa _1 & \kappa _2 & k \\
l_i & l_j & \kappa _{12}
\end{array}
\right\} \times \\
\times \left\{
\begin{array}{ccc}
s & s & \sigma _{12}^{\prime } \\
\sigma _1 & \sigma _2 & k \\
s & s & \sigma _{12}
\end{array}
\right\} B_3
\end{array}
\end{equation}
whereas the second (\ref{eq:gc}):

\begin{equation}
\label{eq:gr-i}
\begin{array}[b]{c}
G_{II}(iji^{\prime }j^{\prime })=B_2+B_4= \\
=\frac 12
\displaystyle {\sum_p}\left( -1\right) ^{k-p}\left( n_i\lambda _in_j\lambda
_j||g^{\left( \kappa _1\kappa _2k,\sigma _1\sigma _2k\right) }||n_i^{\prime
}\lambda _i^{\prime }n_j^{\prime }\lambda _j^{\prime }\right) \times \\
\times \{\left[ \kappa _1,\kappa _2,\sigma _1,\sigma _2\right]
^{-1/2}B_5-\left( -1\right) ^{l_i+l_j^{\prime }}\left\{
\begin{array}{ccc}
\kappa _1 & \kappa _2 & k \\
l_j^{\prime } & l_i & l_j
\end{array}
\right\} \times \\
\times \left\{
\begin{array}{ccc}
\sigma _1 & \sigma _2 & k \\
s & s & s
\end{array}
\right\} \delta \left( n_jl_j,n_i^{\prime }l_i^{\prime }\right) B_6\}.
\end{array}
\end{equation}

We emphasize here that the winding line of interaction in the
Feynman-Goldstone diagram corresponds to the operators of second
quantization in the normal order (Figure 2, $B_1$). Whereas the dotted
interaction line indicates that the second quantization operators are
ordered in pairs of creation-annihilation. In the latter case first comes
the pair on the left side of a Feynman-Goldstone diagram (Figure 2, $B_2$).
Such a notation of two kinds for an interaction line is meaningful only for
two-particle (or more) operators, since for any one particle operator both
the winding and dotted lines correspond to the same order of creation and
annihilation operators.

From expressions (\ref{eq:gr-h}), (\ref{eq:gr-i}) we see that the
two-particle operator in the first form is represented by one
Feynman-Goldstone diagram $B_1$, whereas in the second - by two diagrams $B_2
$ and $B_4$. The diagrams, corresponding to tensorial product, have the
following algebraic expressions:

\begin{equation}
\label{eq:gr-j}B_3=\left[ \left[ a^{\left( \lambda _i\right) }\times
a^{\left( \lambda _j\right) }\right] ^{\left( \kappa _{12}\sigma
_{12}\right) }\times \left[ \stackrel{\sim }{a}^{\left( \lambda _i^{\prime
}\right) }\times \stackrel{\sim }{a}^{\left( \lambda _j^{\prime }\right)
}\right] ^{\left( \kappa _{12}^{\prime }\sigma _{12}^{\prime }\right)
}\right] _{p-p}^{\left( kk\right) },
\end{equation}

\begin{equation}
\label{eq:gr-k}B_5=\left[ \left[ a^{\left( \lambda _i\right) }\times
\stackrel{\sim }{a}^{\left( \lambda _i^{\prime }\right) }\right] ^{\left(
\kappa _1\sigma _1\right) }\times \left[ a^{\left( \lambda _j\right) }\times
\stackrel{\sim }{a}^{\left( \lambda _j^{\prime }\right) }\right] ^{\left(
\kappa _2\sigma _2\right) }\right] _{p-p}^{\left( kk\right) }
\end{equation}

\begin{equation}
\label{eq:gr-l}B_6=\left[ a^{\left( \lambda _i\right) }\times \stackrel{\sim
}{a}^{\left( \lambda _j^{\prime }\right) }\right] _{p-p}^{\left( kk\right) }
\end{equation}

Thus, the obtaining of algebraic expressions from the diagrams $B_3$, $B_5$
and $B_6$ is similar to the case of the diagram $A_2$. The positions of the
second quantization operators in the diagram define their order in tensorial
product: the first place in tensorial product occupies the upper right
second quantization operator, the second - lower right, after them the upper
left and lower left operators follow. The angular momenta diagram defines
their coupling scheme into tensorial product.

Thus, obeying these rules it is possible to easily find the algebraic
counterparts of the diagrams, not forgetting that the arrangement of the
operators must not contradict to their coupling order, i.e. only
neighbouring second quantization operators are coupled into tensorial
product and their disposition order corresponds to coupling scheme.
Otherwise some graphical operations are necessary. Let us present the
simplest of them below as the example for the case, when we have to change
the disposition of the second quantization operators and coupling scheme in
the tensorial product.

%
%
%
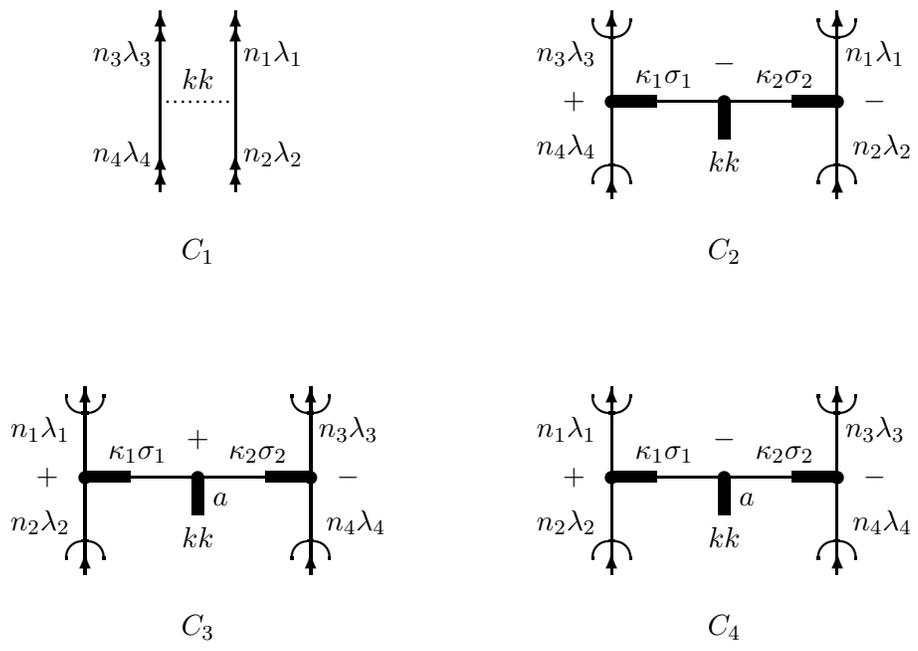
\begin{figure}
\setlength{\unitlength}{1mm}
\begin{picture}(148,105)
\thicklines
\put(20,90){\vector(0,1){10}}
\put(20,100){\vector(0,1){2}}
\put(15,98){\makebox(0,0)[t]{$ n_{3}\lambda_{3}$}}
\put(30,90){\vector(0,1){10}}
\put(30,100){\vector(0,1){2}}
\put(35,98){\makebox(0,0)[t]{$ n_{1}\lambda_{1}$}}
\multiput(20,90)(1,0){10}{\circle*{0.02}}
\put(25,93){\makebox(0,0){$kk$}}
\put(20,80){\line(0,1){10}}
\put(20,80){\vector(0,1){3}}
\put(15,83){\makebox(0,0){$ n_{4}\lambda_{4}$}}
\put(20,78){\vector(0,1){3}}
\put(30,80){\line(0,1){10}}
\put(30,80){\vector(0,1){3}}
\put(35,83){\makebox(0,0){$ n_{2}\lambda_{2}$}}
\put(30,78){\vector(0,1){3}}
\put(25,70){\makebox(0,0){$ C_{1}$}}
%
\put(80,90){\vector(0,1){12}}
\put(80,101){\oval(5,5)[b]}
\put(80,90){\circle*{1,7}}
\put(74,98){\makebox(0,0)[t]{$ n_{3}\lambda_{3}$}}
\put(110,90){\vector(0,1){12}}
\put(110,101){\oval(5,5)[b]}
\put(110,90){\circle*{1,7}}
\put(115,98){\makebox(0,0)[t]{$ n_{1}\lambda_{1}$}}
\put(80,90.6){\line(6,0){6}}
\put(80,90.3){\line(6,0){6}}
\put(80,90){\line(20,0){30}}
\put(80,89.7){\line(6,0){6}}
\put(80,89.4){\line(6,0){6}}
\put(87,93){\makebox(0,0){$\kappa_{1} \sigma_{1}$}}
\put(104,90.6){\line(6,0){6}}
\put(104,90.3){\line(6,0){6}}
\put(104,89.7){\line(6,0){6}}
\put(104,89.4){\line(6,0){6}}
\put(103,93){\makebox(0,0){$\kappa_{2} \sigma_{2}$}}
\put(95,90){\circle*{1,7}}
\put(95,95){\makebox(0,0){${-}$}}
\put(95.6,85){\line(0,1){5}}
\put(95.3,85){\line(0,1){5}}
\put(95,85){\line(0,1){5}}
\put(94.7,85){\line(0,1){5}}
\put(94.4,85){\line(0,1){5}}
\put(95,82){\makebox(0,0){$k k$}}
\put(80,80){\line(0,1){10}}
\put(74,84){\makebox(0,0){$ n_{4}\lambda_{4}$}}
\put(75,90){\makebox(0,0){${+}$}}
\put(80,77){\vector(0,1){3}}
\put(80,79){\oval(5,5)[t]}
\put(110,80){\line(0,1){10}}
\put(116,84){\makebox(0,0){$ n_{2}\lambda_{2}$}}
\put(115,90){\makebox(0,0){${-}$}}
\put(110,77){\vector(0,1){3}}
\put(110,79){\oval(5,5)[t]}
\put(95,70){\makebox(0,0){$ C_{2}$}}
%
\put(10,40){\vector(0,1){12}}
\put(10,51){\oval(5,5)[b]}
\put(10,40){\circle*{1,7}}
\put(4,48){\makebox(0,0)[t]{$ n_{1}\lambda_{1}$}}
\put(40,40){\vector(0,1){12}}
\put(40,51){\oval(5,5)[b]}
\put(40,40){\circle*{1,7}}
\put(45,48){\makebox(0,0)[t]{$ n_{3}\lambda_{3}$}}
\put(10,40.6){\line(6,0){6}}
\put(10,40.3){\line(6,0){6}}
\put(10,40){\line(20,0){30}}
\put(10,39.7){\line(6,0){6}}
\put(10,39.4){\line(6,0){6}}
\put(17,43){\makebox(0,0){$\kappa_{1} \sigma_{1}$}}
\put(34,40.6){\line(6,0){6}}
\put(34,40.3){\line(6,0){6}}
\put(34,39.7){\line(6,0){6}}
\put(34,39.4){\line(6,0){6}}
\put(33,43){\makebox(0,0){$\kappa_{2} \sigma_{2}$}}
\put(25,40){\circle*{1,7}}
\put(25,45){\makebox(0,0){${+}$}}
\put(28,37){\makebox(0,0){${a}$}}
\put(25.6,35){\line(0,1){5}}
\put(25.3,35){\line(0,1){5}}
\put(25,35){\line(0,1){5}}
\put(24.7,35){\line(0,1){5}}
\put(24.4,35){\line(0,1){5}}
\put(25,32){\makebox(0,0){$k k$}}
\put(10,30){\line(0,1){10}}
\put(4,34){\makebox(0,0){$ n_{2}\lambda_{2}$}}
\put(5,40){\makebox(0,0){${+}$}}
\put(10,27){\vector(0,1){3}}
\put(10,29){\oval(5,5)[t]}
\put(40,30){\line(0,1){10}}
\put(46,34){\makebox(0,0){$ n_{4}\lambda_{4}$}}
\put(45,40){\makebox(0,0){${-}$}}
\put(40,27){\vector(0,1){3}}
\put(40,29){\oval(5,5)[t]}
\put(25,20){\makebox(0,0){$ C_{3}$}}
%
\put(80,40){\vector(0,1){12}}
\put(80,51){\oval(5,5)[b]}
\put(80,40){\circle*{1,7}}
\put(74,48){\makebox(0,0)[t]{$ n_{1}\lambda_{1}$}}
\put(110,40){\vector(0,1){12}}
\put(110,51){\oval(5,5)[b]}
\put(110,40){\circle*{1,7}}
\put(115,48){\makebox(0,0)[t]{$ n_{3}\lambda_{3}$}}
\put(80,40.6){\line(6,0){6}}
\put(80,40.3){\line(6,0){6}}
\put(80,40){\line(20,0){30}}
\put(80,39.7){\line(6,0){6}}
\put(80,39.4){\line(6,0){6}}
\put(87,43){\makebox(0,0){$\kappa_{1} \sigma_{1}$}}
\put(104,40.6){\line(6,0){6}}
\put(104,40.3){\line(6,0){6}}
\put(104,39.7){\line(6,0){6}}
\put(104,39.4){\line(6,0){6}}
\put(103,43){\makebox(0,0){$\kappa_{2} \sigma_{2}$}}
\put(95,40){\circle*{1,7}}
\put(95,45){\makebox(0,0){${-}$}}
\put(98,37){\makebox(0,0){${a}$}}
\put(95.6,35){\line(0,1){5}}
\put(95.3,35){\line(0,1){5}}
\put(95,35){\line(0,1){5}}
\put(94.7,35){\line(0,1){5}}
\put(94.4,35){\line(0,1){5}}
\put(95,32){\makebox(0,0){$k k$}}
\put(80,30){\line(0,1){10}}
\put(74,34){\makebox(0,0){$ n_{2}\lambda_{2}$}}
\put(75,40){\makebox(0,0){${+}$}}
\put(80,27){\vector(0,1){3}}
\put(80,29){\oval(5,5)[t]}
\put(110,30){\line(0,1){10}}
\put(116,34){\makebox(0,0){$ n_{4}\lambda_{4}$}}
\put(115,40){\makebox(0,0){${-}$}}
\put(110,27){\vector(0,1){3}}
\put(110,29){\oval(5,5)[t]}
\put(95,20){\makebox(0,0){$ C_{4}$}}
\end{picture}
\caption{Diagrams for graphical transformations.}
\label{cc}
\end{figure}
%
%

Suppose, we have the following correspondence between diagrams (Figure 3):

\begin{equation}
\label{eq:grf-a}C_1\longrightarrow C_2,
\end{equation}
in which the second quantization operators are in the order $a^{(\lambda
_3)} $ $\tilde a^{(\lambda _4)}$ $a^{(\lambda _1)}$ $\tilde a^{(\lambda _2)}$%
. Our goal is to obtain the diagram corresponding to the order $a^{(\lambda
_1)}$ $\tilde a^{(\lambda _2)}$ $a^{(\lambda _3)}$ $\tilde a^{(\lambda _4)}$%
. Bearing in mind that the second quantization operators anticommute with
each other and they all act on different electronic shells and we are not
changing the order of their coupling into tensorial product, we arrive at

\begin{equation}
\label{eq:grf-b}C_1\longrightarrow (-1)^4C_3=C_3.
\end{equation}

Let us also discuss another situation: we have defined the disposition of
the operators and we want to couple them into certain tensorial product.
Suppose that we want to represent graphically the following tensorial
product:

\begin{equation}
\label{eq:grf-c}\left[ \left[ a^{(\lambda _1)}\times \tilde a^{(\lambda
_2)}\right] ^{\left( \kappa _1\sigma _1\right) }\times \left[ a^{(\lambda
_3)}\times \tilde a^{(\lambda _4)}\right] ^{\left( \kappa _2\sigma _2\right)
}\right] ^{\left( \kappa \sigma \right) }.
\end{equation}

For this purpose we have to rearrange the generalized Clebsch-Gordan
coefficient, which is defining the scheme of coupling of the operators into
the tensorial product. It is easy to notice that this coefficient will
properly define the tensorial product, if we change the sign of the vertex ''%
$a$'' in diagram $C_3$. Making use of (\ref{eq:gr-da}) we find:

\begin{equation}
\label{eq:grf-d}C_1\longrightarrow (-1)^{\kappa _1+\kappa _2-\kappa +\sigma
_1+\sigma _2-\sigma }C_4.
\end{equation}

The procedures described are fairly simple, however, they are sufficient for
the majority of cases. The more complete description of this generalized
graphical approach may be found in Gaigalas {\it et al} 1985, Gaigalas 1985,
Gaigalas and Merkelis 1987.

\section{\bf Quasispin Formalism}

A wave function with $u$ shells in $LS$ coupling may be denoted in the form

\begin{equation}
\label{eq:gh}
\begin{array}{c}
\psi _u\left( LSM_LM_S\right) \equiv \\
\equiv |n_1l_1^{N_1}n_2l_2^{N_2}...n_ul_u^{N_u}\alpha _1L_1S_1\alpha
_2L_2S_2...\alpha _uL_uS_uALSM_LM_S),
\end{array}
\end{equation}
where $A$ stands for all intermediate quantum numbers, depending on the
order of coupling of momenta $\alpha _iL_iS_i$.

As we shall see later on, it is very convenient for the calculations of
matrix elements to use quasispin formalism. Then $a_{m_\lambda }^{\left(
\lambda \right) }$ and $\stackrel{\sim }{a}_{m_\lambda }^{\left( \lambda
\right) }$ are components of the tensor $a_{m_qm_\lambda }^{\left( q\lambda
\right) }$ , having in additional quasispin space the rank $q=\frac 12$ and
projections $m_q=\pm \frac 12$, i.e. $a_{\frac 12m_\lambda }^{\left(
q\lambda \right) }=a_{m_lm_s}^{\left( ls\right) }$ and $a_{-\frac
12m_\lambda }^{\left( q\lambda \right) }=\stackrel{\sim }{a}%
_{m_lm_s}^{\left( ls\right) }$.

In the quasispin representation, for a wave function of the shell of
equivalent electrons $|nl^N\alpha LS)$ a label $Q$ - quasispin momentum of
the shell - is introduced, which is related to the seniority quantum number $%
\nu $, namely, $Q=\left( 2l+1-\nu \right) /2$, and its projection, $%
M_Q=\left( N-2l-1\right) /2$. Here $\alpha $ denotes all additional quantum
numbers needed for the one-to-one classification of the energy levels. Then,
we can rewrite the wave function (\ref{eq:gh})

\begin{equation}
\label{eq:gha}
\begin{array}[b]{c}
\psi _u\left( LSM_LM_S\right) \equiv \\
\equiv |n_1l_1n_2l_2...n_ul_u\alpha _1L_1S_1Q_1M_{Q_1}\alpha
_2L_2S_2Q_2M_{Q_2}... \\
\alpha _uL_uS_uQ_uM_{Q_u}ALSM_LM_S).
\end{array}
\end{equation}

Making use of the Wigner-Eckart theorem in quasispin space of a shell $l^N$,

\begin{equation}
\label{eq:gghhaa}
\begin{array}[b]{c}
\left( l\;\alpha QLSM_Q||T_{m_q}^{\left( qls\right) }||l\;\alpha ^{\prime
}Q^{\prime }L^{\prime }S^{\prime }M_Q^{\prime }\right) =\left( -1\right)
^{2q}\left[ Q\right] ^{-1/2}\times \\
\times \left[
\begin{array}{ccc}
Q^{\prime } & q & Q \\
M_Q^{\prime } & m_q & M_Q
\end{array}
\right] \left( l\;\alpha QLS|||T^{\left( qls\right) }|||l\;\alpha ^{\prime
}Q^{\prime }L^{\prime }S^{\prime }\right)
\end{array}
\end{equation}
it is possible to define the notions of a completely reduced matrix element $%
\left( l\;\alpha QLS|||T^{\left( qls\right) }|||l\;\alpha ^{\prime
}Q^{\prime }L^{\prime }S^{\prime }\right) $ and subcoefficient of fractional
parentage (reduced coefficient of fractional parentage) $\left( l\;\alpha
QLS|||a^{\left( qls\right) }|||l\;\alpha ^{\prime }Q^{\prime }L^{\prime
}S^{\prime }\right) $. In (\ref{eq:gghhaa}) $T_{m_q}^{\left( qls\right) }$
is any tensor with rank $q$ and its projection $m_q$ in quasispin space and
on the right-hand side of this equation only the Clebsch-Gordan coefficient $%
\left[
\begin{array}{ccc}
Q^{\prime } & q & Q \\
M_Q^{\prime } & m_q & M_Q
\end{array}
\right] $ depends on the number $N$ of equivalent electrons.

According to Rudzikas and Kaniauskas 1984 we have the following relation
between the coefficients of fractional parentage and completely reduced
matrix elements $\left( l\;\alpha QLS|||a^{\left( qls\right) }|||l\;\alpha
^{\prime }Q^{\prime }L^{\prime }S^{\prime }\right) $ of the operator of
second quantization $a^{\left( qls\right) }$:

\begin{equation}
\label{eq:gi}
\begin{array}[b]{c}
\left( l^N\;\alpha QLS||l^{N-1}\;\left( \alpha ^{\prime }Q^{\prime
}L^{\prime }S^{\prime }\right) l\right) =\left( -1\right) ^{N-1}\left(
N\left[ Q,L,S\right] \right) ^{-1/2}\times \\
\times \left[
\begin{array}{ccc}
Q^{\prime } & 1/2 & Q \\
M_Q^{\prime } & 1/2 & M_Q
\end{array}
\right] \left( l\;\alpha QLS|||a^{\left( qls\right) }|||l\;\alpha ^{\prime
}Q^{\prime }L^{\prime }S^{\prime }\right) .
\end{array}
\end{equation}

Tables of numerical values of $\left( l\;\alpha QLS|||a^{\left( qls\right)
}|||l\;\alpha Q^{\prime }L^{\prime }S^{\prime }\right) $ are presented in
Rudzikas and Kaniauskas 1984 when $l=0,1,2$. For the tensorial product of
two one-electron operators, the submatrix element equals

\begin{equation}
\label{eq:gj}
\begin{array}[b]{c}
\left( nl^N\;\alpha QLS||\left[ a_{m_{q1}}^{\left( q\lambda \right) }\times
a_{m_{q2}}^{\left( q\lambda \right) }\right] ^{\left( k_1k_2\right)
}||nl^{N^{\prime }}\;\alpha ^{\prime }Q^{\prime }L^{\prime }S^{\prime
}\right) = \\
=\displaystyle {\sum_{\epsilon ,m_\epsilon }}\left[ Q\right] ^{-1/2}\left[
\begin{array}{ccc}
q & q & \epsilon \\
m_{q1} & m_{q2} & m_\epsilon
\end{array}
\right] \left[
\begin{array}{ccc}
Q^{\prime } & \epsilon & Q \\
M_Q^{\prime } & m_\epsilon & M_Q
\end{array}
\right] \times \\
\times \left( nl\;\alpha QLS|||W^{\left( \epsilon k_1k_2\right)
}|||nl\;\alpha ^{\prime }Q^{\prime }L^{\prime }S^{\prime }\right) .
\end{array}
\end{equation}

On the right-hand side of equations (\ref{eq:gi}) and (\ref{eq:gj}) only the
Clebsch-Gordan coefficient $\left[
\begin{array}{ccc}
Q^{\prime } & \epsilon & Q \\
M_Q^{\prime } & m_\epsilon & M_Q
\end{array}
\right] $ depends on the number $N$ of equivalent electrons.

$\left( nl\;\alpha QLS|||W^{\left( \epsilon k_1k_2\right) }|||nl\;\alpha
^{\prime }Q^{\prime }L^{\prime }S^{\prime }\right) $ denotes reduced in
quasispin space submatrix element (completely reduced matrix element) of the
triple tensor $W^{\left( \epsilon k_1k_2\right) }\left( nl,nl\right) =\left[
a^{\left( qls\right) }\times a^{\left( qls\right) }\right] ^{\left( \epsilon
k_1k_2\right) }$. It is related to the completely reduced coefficients
(subcoefficients) of fractional parentage in a following way:

\begin{equation}
\label{eq:gk}
\begin{array}[b]{c}
\left( nl\;\alpha QLS|||W^{\left( \epsilon k_1k_2\right) }|||nl\;\alpha
^{\prime }Q^{\prime }L^{\prime }S^{\prime }\right) = \\
=\left( -1\right) ^{Q+L+S+Q^{\prime }+L^{\prime }+S^{\prime }+\epsilon
+k_1+k_2}\left[ \epsilon ,k_1,k_2\right] ^{1/2}\times \\
\times
\displaystyle {\sum_{\alpha ^{\prime \prime }Q^{\prime \prime }L^{\prime
\prime }S^{\prime \prime }}}\left( l\;\alpha QLS|||a^{\left( qls\right)
}|||l\;\alpha ^{\prime \prime }Q^{\prime \prime }L^{\prime \prime }S^{\prime
\prime }\right) \times \\ \times \left( l\;\alpha ^{\prime \prime }Q^{\prime
\prime }L^{\prime \prime }S^{\prime \prime }|||a^{\left( qls\right)
}|||l\;\alpha ^{\prime }Q^{\prime }L^{\prime }S^{\prime }\right) \times \\
\times \left\{
\begin{array}{ccc}
q & q & \epsilon \\
Q^{\prime } & Q & Q^{\prime \prime }
\end{array}
\right\} \left\{
\begin{array}{ccc}
l & l & k_1 \\
L^{\prime } & L & L^{\prime \prime }
\end{array}
\right\} \left\{
\begin{array}{ccc}
s & s & k_2 \\
S^{\prime } & S & S^{\prime \prime }
\end{array}
\right\} .
\end{array}
\end{equation}

So, by applying the quasispin method for calculating the matrix elements of
any operator, we can use the reduced coefficients of fractional parentage or
the tensors (for example $W^{\left( \epsilon k_1k_2\right) }\left(
nl,nl\right) $), which are independent of the occupation number of the shell
for a given $\nu $. The main advantage of this approach is that the standard
data tables in such a case will be much smaller in comparison with tables of
the usual coefficients and, therefore, many summations will be less
time-consuming. Also one can see that in such an approach the submatrix
elements of standard tensors and subcoefficients of fractional parentage
actually can be treated in a uniform way as they all are the completely
reduced matrix elements of the second quantization operators. Hence, all
methodology of calculation of matrix elements will be much more universal in
comparison with the traditional one.

\section{\bf Matrix Elements in the Case of\ Two\ Open\ Shells\ of\
Equivalent\ Electrons}

The aim of this section is to illustrate the usage of abovementioned
methodology to obtain the expressions for matrix elements of a two-particle
operator, when the wave function (\ref{eq:gha}) has two open shells $%
n_1l_1^{N_1}$ $n_2l_2^{N_2}$. Then it may be written as

\begin{equation}
\label{eq:gl}\psi _2\left( LSM_LM_S\right) \equiv |n_1l_1n_2l_2\alpha
_1L_1S_1Q_1M_{Q_1}\alpha _2L_2S_2Q_2M_{Q_2}LSM_LM_S).
\end{equation}

To find numerical value of physical quantity of two-electron operator one
ought to have the expressions for its matrix elements within each shell of
equivalent electrons and between each pair of the shells, including
non-diagonal, with respect to configurations, matrix elements.

While calculating the diagonal matrix elements between functions (\ref{eq:gl}%
), the quantum numbers $n_i\lambda _i,n_i^{\prime }\lambda _i^{\prime
},n_j\lambda _j,n_j^{\prime }\lambda _j^{\prime }$ in two alternative
expressions (\ref{eq:gb}) (\ref{eq:gc}) acquire the following values:

1. $n_i\lambda _i=n_i^{\prime }\lambda _i^{\prime }=n_j\lambda
_j=n_j^{\prime }\lambda _j^{\prime }=n_1l_1s.$ (All the operators of second
quantization act upon the first shell).

2. $n_i\lambda _i=n_i^{\prime }\lambda _i^{\prime }=n_j\lambda
_j=n_j^{\prime }\lambda _j^{\prime }=n_2l_2s.$ (All the operators of second
quantization act upon the second shell).

3. $n_i\lambda _i=n_i^{\prime }\lambda _i^{\prime }=n_1l_1s,\;n_j\lambda
_j=n_j^{\prime }\lambda _j^{\prime }=n_2l_2s.$

4. $n_j\lambda _j=n_j^{\prime }\lambda _j^{\prime }=n_1l_1s,\;n_i\lambda
_i=n_i^{\prime }\lambda _i^{\prime }=n_2l_2s.$

5. $n_i\lambda _i=n_j^{\prime }\lambda _j^{\prime }=n_1l_1s,\;n_i^{\prime
}\lambda _i^{\prime }=n_j\lambda _j=n_2l_2s.$

6. $n_i^{\prime }\lambda _i^{\prime }=n_j\lambda _j=n_1l_1s,\;n_i\lambda
_i=n_j^{\prime }\lambda _j^{\prime }=n_2l_2s.$

In the first case the matrix elements of operator in the first (using (\ref
{eq:gb})) and the second (using (\ref{eq:gc})) forms are equal respectively

\begin{equation}
\label{eq:gm}
\begin{array}[b]{c}
(n_1l_1^{N_1}n_2l_2^{N_2}\alpha _1L_1S_1Q_1M_{Q_1}\alpha
_2L_2S_2Q_2M_{Q_2}LSM_LM_S|G_I(1111) \\
|n_1l_1^{N_1}n_2l_2^{N_2}\alpha _1^{\prime }L_1^{\prime }S_1^{\prime
}Q_1^{\prime }M_{Q_1}\alpha _2^{\prime }L_2^{\prime }S_2^{\prime
}Q_2^{\prime }M_{Q_2}L^{\prime }S^{\prime }M_L^{\prime }M_S^{\prime })= \\
\\
=-\frac 12
\displaystyle {\sum_{\kappa _{12}\kappa _{12}^{^{\prime }}\sigma _{12}\sigma
_{12}^{^{\prime }}}}\displaystyle {\sum_p}\left( -1\right) ^{k-p}\left[
\kappa _1,\kappa _2,\sigma _1,\sigma _2\right] ^{1/2}\times \\ \times \left(
n_1\lambda _1n_1\lambda _1||g^{\left( \kappa _1\kappa _2k,\sigma _1\sigma
_2k\right) }||n_1\lambda _1n_1\lambda _1\right) \times \\
\times \left\{
\begin{array}{ccc}
l_1 & l_1 & \kappa _{12}^{\prime } \\
\kappa _1 & \kappa _2 & k \\
l_1 & l_1 & \kappa _{12}
\end{array}
\right\} \left\{
\begin{array}{ccc}
s & s & \sigma _{12}^{\prime } \\
\sigma _1 & \sigma _2 & k \\
s & s & \sigma _{12}
\end{array}
\right\} \times \\
\times (n_1l_1^{N_1}n_2l_2^{N_2}\alpha _1L_1S_1Q_1M_{Q_1}\alpha
_2L_2S_2Q_2M_{Q_2}LSM_LM_S| \\
\left[ \left[ a^{\left( \lambda _1\right) }\times a^{\left( \lambda
_1\right) }\right] ^{\left( \kappa _{12}\sigma _{12}\right) }\times \left[
\stackrel{\sim }{a}^{\left( \lambda _1\right) }\times \stackrel{\sim }{a}%
^{\left( \lambda _1\right) }\right] ^{\left( \kappa _{12}^{\prime }\sigma
_{12}^{\prime }\right) }\right] _{p-p}^{\left( kk\right) } \\
|n_1l_1^{N_1}n_2l_2^{N_2}\alpha _1^{\prime }L_1^{\prime }S_1^{\prime
}Q_1^{\prime }M_{Q_1}\alpha _2^{\prime }L_2^{\prime }S_2^{\prime
}Q_2^{\prime }M_{Q_2}L^{\prime }S^{\prime }M_L^{\prime }M_S^{\prime }),
\end{array}
\end{equation}

\begin{equation}
\label{eq:gn}
\begin{array}[b]{c}
(n_1l_1^{N_1}n_2l_2^{N_2}\alpha _1L_1S_1Q_1M_{Q_1}\alpha
_2L_2S_2Q_2M_{Q_2}LSM_LM_S|G_{II}(1111) \\
|n_1l_1^{N_1}n_2l_2^{N_2}\alpha _1^{\prime }L_1^{\prime }S_1^{\prime
}Q_1^{\prime }M_{Q_1}\alpha _2^{\prime }L_2^{\prime }S_2^{\prime
}Q_2^{\prime }M_{Q_2}L^{\prime }S^{\prime }M_L^{\prime }M_S^{\prime })= \\
=\frac 12
\displaystyle {\sum_p}\left( -1\right) ^{k-p}\left( n_1\lambda _1n_1\lambda
_1||g^{\left( \kappa _1\kappa _2k,\sigma _1\sigma _2k\right) }||n_1\lambda
_1n_1\lambda _1\right) \times \\ \times \{\left[ \kappa _1,\kappa _2,\sigma
_1,\sigma _2\right] ^{-1/2}\times \\
\times (n_1l_1^{N_1}n_2l_2^{N_2}\alpha _1L_1S_1Q_1M_{Q_1}\alpha
_2L_2S_2Q_2M_{Q_2}LSM_LM_S| \\
\left[ \left[ a^{\left( \lambda _1\right) }\times
\stackrel{\sim }{a}^{\left( \lambda _1\right) }\right] ^{\left( \kappa
_1\sigma _1\right) }\times \left[ a^{\left( \lambda _1\right) }\times
\stackrel{\sim }{a}^{\left( \lambda _1\right) }\right] ^{\left( \kappa
_2\sigma _2\right) }\right] _{p-p}^{\left( kk\right) } \\
|n_1l_1^{N_1}n_2l_2^{N_2}\alpha _1^{\prime }L_1^{\prime }S_1^{\prime
}Q_1^{\prime }M_{Q_1}\alpha _2^{\prime }L_2^{\prime }S_2^{\prime
}Q_2^{\prime }M_{Q_2}L^{\prime }S^{\prime }M_L^{\prime }M_S^{\prime })- \\
-\left\{
\begin{array}{ccc}
\kappa _1 & \kappa _2 & k \\
l_1 & l_1 & l_1
\end{array}
\right\} \left\{
\begin{array}{ccc}
\sigma _1 & \sigma _2 & k \\
s & s & s
\end{array}
\right\} \times \\
\times (n_1l_1^{N_1}n_2l_2^{N_2}\alpha _1L_1S_1Q_1M_{Q_1}\alpha
_2L_2S_2Q_2M_{Q_2}LSM_LM_S| \\
\left[ a^{\left( \lambda _1\right) }\times
\stackrel{\sim }{a}^{\left( \lambda _1\right) }\right] _{p-p}^{\left(
kk\right) } \\ |n_1l_1^{N_1}n_2l_2^{N_2}\alpha _1^{\prime }L_1^{\prime
}S_1^{\prime }Q_1^{\prime }M_{Q_1}\alpha _2^{\prime }L_2^{\prime
}S_2^{\prime }Q_2^{\prime }M_{Q_2}L^{\prime }S^{\prime }M_L^{\prime
}M_S^{\prime })\}.
\end{array}
\end{equation}

Schematically these expressions may be written down as

\begin{equation}
\label{eq:go}
\begin{array}[b]{c}
(n_1l_1^{N_1}n_2l_2^{N_2}\alpha _1L_1S_1Q_1M_{Q_1}\alpha
_2L_2S_2Q_2M_{Q_2}LSM_LM_S|G(1111) \\
|n_1l_1^{N_1}n_2l_2^{N_2}\alpha _1^{\prime }L_1^{\prime }S_1^{\prime
}Q_1^{\prime }M_{Q_1}\alpha _2^{\prime }L_2^{\prime }S_2^{\prime
}Q_2^{\prime }M_{Q_2}L^{\prime }S^{\prime }M_L^{\prime }M_S^{\prime })= \\
=
\displaystyle {\sum_{\kappa _{12},\sigma _{12},\kappa _{12}^{\prime }\sigma
_{12}^{\prime }}}\Theta \left( \kappa _{12},\sigma _{12},\kappa
_{12}^{\prime },\sigma _{12}^{\prime },n_1,\lambda _1\right) \times \\
\times (n_1l_1^{N_1}n_2l_2^{N_2}\alpha _1L_1S_1Q_1M_{Q_1}\alpha
_2L_2S_2Q_2M_{Q_2}LSM_LM_S| \\
A_{p-p}^{\left( kk\right) }\left( \kappa _{12},\sigma _{12},\kappa
_{12}^{\prime },\sigma _{12}^{\prime },n_1,\lambda _1\right) \\
|n_1l_1^{N_1}n_2l_2^{N_2}\alpha _1^{\prime }L_1^{\prime }S_1^{\prime
}Q_1^{\prime }M_{Q_1}\alpha _2^{\prime }L_2^{\prime }S_2^{\prime
}Q_2^{\prime }M_{Q_2}L^{\prime }S^{\prime }M_L^{\prime }M_S^{\prime }),
\end{array}
\end{equation}
where $\Theta \left( \kappa _{12},\sigma _{12},\kappa _{12}^{\prime },\sigma
_{12}^{\prime },n_1,\lambda _1\right) $ is proportional to the radial part
of an operator, and $A_{p-p}^{\left( kk\right) }\left( \kappa _{12},\sigma
_{12},\kappa _{12}^{\prime },\sigma _{12}^{\prime },n_1,\lambda _1\right) $,
- to the spin-angular part of it. In the first form

\begin{equation}
\label{eq:gp}
\begin{array}{c}
A_{p-p}^{\left( kk\right) }\left( \kappa _{12},\sigma _{12},\kappa
_{12}^{\prime },\sigma _{12}^{\prime },n_1,\lambda _1\right) = \\
=\left[ \left[ a^{\left( \lambda _1\right) }\times a^{\left( \lambda
_1\right) }\right] ^{\left( \kappa _{12}\sigma _{12}\right) }\times \left[
\stackrel{\sim }{a}^{\left( \lambda _1\right) }\times \stackrel{\sim }{a}%
^{\left( \lambda _1\right) }\right] ^{\left( \kappa _{12}^{\prime }\sigma
_{12}^{\prime }\right) }\right] _{p-p}^{\left( kk\right) }
\end{array}
\end{equation}
whereas in the second form ($\kappa _{12}=\kappa _1,\sigma _{12}=\sigma
_1,\kappa _{12}^{\prime }=\kappa _2,\sigma _{12}^{\prime }=\sigma _2$)

\begin{equation}
\label{eq:gr}
\begin{array}[b]{c}
A_{p-p}^{\left( kk\right) }\left( \kappa _1,\sigma _1,\kappa _2,\sigma
_2,n_1,\lambda _1\right) =\{\left[ \kappa _1,\kappa _2,\sigma _1,\sigma
_2\right] ^{-1/2}\times \\
\times \left[ \left[ a^{\left( \lambda _1\right) }\times
\stackrel{\sim }{a}^{\left( \lambda _1\right) }\right] ^{\left( \kappa
_1\sigma _1\right) }\times \left[ a^{\left( \lambda _1\right) }\times
\stackrel{\sim }{a}^{\left( \lambda _1\right) }\right] ^{\left( \kappa
_2\sigma _2\right) }\right] _{p-p}^{\left( kk\right) }- \\ -\left\{
\begin{array}{ccc}
\kappa _1 & \kappa _2 & k \\
l_1 & l_1 & l_1
\end{array}
\right\} \left\{
\begin{array}{ccc}
\sigma _1 & \sigma _2 & k \\
s & s & s
\end{array}
\right\} \left[ a^{\left( \lambda _{i1}\right) }\times \stackrel{\sim }{a}%
^{\left( \lambda _1\right) }\right] _{p-p}^{\left( kk\right) }\}.
\end{array}
\end{equation}

So, in order to calculate the spin-angular parts of matrix elements of
operators (\ref{eq:gb}), (\ref{eq:gc}), we have to obtain at first the
matrix elements of operators $A_{p-p}^{\left( kk\right) }\left( \kappa
_{12},\sigma _{12},\kappa _{12}^{\prime },\sigma _{12}^{\prime },n_1,\lambda
_1\right) $. By using the Wigner-Eckart theorem, we find:

\begin{equation}
\label{eq:gs}
\begin{array}[b]{c}
(n_1l_1^{N_1}n_2l_2^{N_2}\alpha _1L_1S_1Q_1M_{Q_1}\alpha
_2L_2S_2Q_2M_{Q_2}LSM_LM_S| \\
A_{p-p}^{\left( kk\right) }\left( \kappa _{12},\sigma _{12},\kappa
_{12}^{\prime },\sigma _{12}^{\prime },n_1,\lambda _1\right) \\
|n_1l_1^{N_1}n_2l_2^{N_2}\alpha _1^{\prime }L_1^{\prime }S_1^{\prime
}Q_1^{\prime }M_{Q_1}\alpha _2^{\prime }L_2^{\prime }S_2^{\prime
}Q_2^{\prime }M_{Q_2}L^{\prime }S^{\prime }M_L^{\prime }M_S^{\prime })= \\
=\left[ L,S\right] ^{-1/2}\left[
\begin{array}{ccc}
L^{\prime } & k & L \\
M_{L^{\prime }} & p & M_L
\end{array}
\right] \left[
\begin{array}{ccc}
S^{\prime } & k & S \\
M_{S^{\prime }} & -p & M_S
\end{array}
\right] \times \\
\times (n_1l_1^{N_1}n_2l_2^{N_2}\alpha _1L_1S_1Q_1M_{Q_1}\alpha
_2L_2S_2Q_2M_{Q_2}LS|| \\
A^{\left( kk\right) }\left( \kappa _{12},\sigma _{12},\kappa _{12}^{\prime
},\sigma _{12}^{\prime },n_1,\lambda _1\right) \\
||n_1l_1^{N_1}n_2l_2^{N_2}\alpha _1^{\prime }L_1^{\prime }S_1^{\prime
}Q_1^{\prime }M_{Q_1}\alpha _2^{\prime }L_2^{\prime }S_2^{\prime
}Q_2^{\prime }M_{Q_2}L^{\prime }S^{\prime }).
\end{array}
\end{equation}

Then we proceed with analyzing the submatrix elements. As the operator $%
A^{\left( kk\right) }\left( \kappa _{12},\sigma _{12},\kappa _{12}^{\prime
},\sigma _{12}^{\prime },n_1,\lambda _1\right) $ acts here upon the first
shell only, then, using the expression (4.7) from Jucys and Savukynas 1973,
namely,

\begin{equation}
\label{eq:js-a}
\begin{array}[b]{c}
(\alpha _1j_1\alpha _2j_2j||A_1^{\left( k\right) }||\alpha _1j_1\alpha
_2j_2j)=\delta \left( \alpha _2j_2,\alpha _2^{\prime }j_2^{\prime }\right)
\left( -1\right) ^{j_1+j_2+j^{\prime }+k}\times \\
\times \left[ j,j^{\prime }\right] ^{1/2}(\alpha _1j_1||A_1^{\left( k\right)
}||\alpha _1j_1)\left\{
\begin{array}{ccc}
j_1 & j & j_2 \\
j^{\prime } & j_1^{\prime } & k
\end{array}
\right\} ,
\end{array}
\end{equation}
we obtain:

\begin{equation}
\label{eq:gt}
\begin{array}[b]{c}
(n_1l_1^{N_1}n_2l_2^{N_2}\alpha _1L_1S_1Q_1M_{Q_1}\alpha
_2L_2S_2Q_2M_{Q_2}LS|| \\
A^{\left( kk\right) }\left( \kappa _{12},\sigma _{12},\kappa _{12}^{\prime
},\sigma _{12}^{\prime },n_1,\lambda _1\right) \\
||n_1l_1^{N_1}n_2l_2^{N_2}\alpha _1^{\prime }L_1^{\prime }S_1^{\prime
}Q_1^{\prime }M_{Q_1}\alpha _2^{\prime }L_2^{\prime }S_2^{\prime
}Q_2^{\prime }M_{Q_2}L^{\prime }S^{\prime })= \\
=\left( -1\right) ^{L_1+S_1+L_2+S_2+L^{\prime }+S^{\prime }+2k}\left[
L,S,L^{\prime },S^{\prime }\right] ^{1/2}\times \\
\times \left\{
\begin{array}{ccc}
L_1 & L & L_2 \\
L^{\prime } & L_1^{\prime } & k
\end{array}
\right\} \left\{
\begin{array}{ccc}
S_1 & S & S_2 \\
S^{\prime } & S_1^{\prime } & k
\end{array}
\right\} \times \\
\times (n_1l_1^{N_1}\;\alpha _1Q_1L_1S_1||A^{\left( kk\right) }\left( \kappa
_{12},\sigma _{12},\kappa _{12}^{\prime },\sigma _{12}^{\prime },n_1,\lambda
_1\right) \\
||n_1l_1^{N_1}\;\alpha _1^{\prime }Q_1^{\prime }L_1^{\prime }S_1^{\prime }).
\end{array}
\end{equation}

Then there remains only to obtain the formulas for the following submatrix
elements:

\begin{equation}
\label{eq:gv}(nl^N\;\alpha QLS||\left[ a^{\left( \lambda \right) }\times
\stackrel{\sim }{a}^{\left( \lambda \right) }\right] ^{\left( kk\right)
}||nl^N\;\alpha ^{\prime }Q^{\prime }L^{\prime }S^{\prime }),
\end{equation}

\begin{equation}
\label{eq:gz}(nl^N\;\alpha QLS||\left[ \left[ a^{\left( \lambda \right)
}\times \stackrel{\sim }{a}^{\left( \lambda \right) }\right] ^{\left( \kappa
_1\sigma _1\right) }\times \left[ a^{\left( \lambda \right) }\times
\stackrel{\sim }{a}^{\left( \lambda \right) }\right] ^{\left( \kappa
_2\sigma _2\right) }\right] ^{\left( kk\right) }||nl^N\;\alpha ^{\prime
}Q^{\prime }L^{\prime }S^{\prime }),
\end{equation}

\begin{equation}
\label{eq:gga}
\begin{array}[b]{c}
(nl^N\;\alpha QLS||\left[ \left[ a^{\left( \lambda \right) }\times a^{\left(
\lambda \right) }\right] ^{\left( \kappa _{12}\sigma _{12}\right) }\times
\left[
\stackrel{\sim }{a}^{\left( \lambda \right) }\times \stackrel{\sim }{a}%
^{\left( \lambda \right) }\right] ^{\left( \kappa _{12}^{\prime }\sigma
_{12}^{\prime }\right) }\right] ^{\left( kk\right) } \\ ||nl^N\;\alpha
^{\prime }Q^{\prime }L^{\prime }S^{\prime }).
\end{array}
\end{equation}

Basing ourselves upon the expressions (\ref{eq:gj}), (\ref{eq:gk}), we
straightforwardly find the value of a submatrix element (\ref{eq:gv}). The
values of submatrix elements (\ref{eq:gz}), (\ref{eq:gga}) follow directly
from the expression (2.28) of Jucys and Savukynas 1973

\begin{equation}
\label{js-b}
\begin{array}[b]{c}
(\alpha j||\left[ A^{\left( k_1\right) }\times B^{\left( k_2\right) }\right]
^{\left( k\right) }||\alpha ^{\prime }j^{\prime })=\left( -1\right)
^{j+j^{\prime }+k}\left[ k\right] ^{1/2}\times \\
\times \displaystyle {\sum_{\alpha ^{\prime \prime }j^{\prime \prime }}}%
(\alpha j||A^{\left( k_1\right) }||\alpha ^{\prime \prime }j^{\prime \prime
})(\alpha ^{\prime \prime }j^{\prime \prime }||B^{\left( k_2\right)
}||\alpha ^{\prime }j^{\prime })\left\{
\begin{array}{ccc}
k_1 & k_2 & k \\
j^{\prime } & j & j^{\prime \prime }
\end{array}
\right\} .
\end{array}
\end{equation}

So we have:

\begin{equation}
\label{eq:ggb}
\begin{array}[b]{c}
(nl^N\;\alpha QLS||\left[ \left[ a_{m_{q1}}^{\left( q\lambda \right) }\times
a_{m_{q2}}^{\left( q\lambda \right) }\right] ^{\left( \kappa _1\sigma
_1\right) }\times \left[ a_{m_{q3}}^{\left( q\lambda \right) }\times
a_{m_{q4}}^{\left( q\lambda \right) }\right] ^{\left( \kappa _2\sigma
_2\right) }\right] ^{\left( kk\right) } \\
||nl^{N^{\prime }}\;\alpha ^{\prime }Q^{\prime }L^{\prime }S^{\prime
})=\left( -1\right) ^{L+S+L^{\prime }+S^{\prime }+2k}\left[ k\right] \times
\\
\times \displaystyle {\sum_{\alpha ^{\prime \prime }Q^{\prime \prime
}L^{\prime \prime }S^{\prime \prime }}}\left\{
\begin{array}{ccc}
\kappa _1 & \kappa _2 & k \\
L^{\prime } & L & L^{\prime \prime }
\end{array}
\right\} \left\{
\begin{array}{ccc}
\sigma _1 & \sigma _2 & k \\
S^{\prime } & S & S^{\prime \prime }
\end{array}
\right\} \times \\
\times (nl^N\;\alpha QLS||\left[ a_{m_{q1}}^{\left( q\lambda \right) }\times
a_{m_{q2}}^{\left( q\lambda \right) }\right] ^{\left( \kappa _1\sigma
_1\right) }||nl^{N^{\prime \prime }}\;\alpha ^{\prime \prime }Q^{\prime
\prime }L^{\prime \prime }S^{\prime \prime })\times \\
\times (nl^{N^{\prime \prime }}\;\alpha ^{\prime \prime }Q^{\prime \prime
}L^{\prime \prime }S^{\prime \prime }||\left[ a_{m_{q3}}^{\left( q\lambda
\right) }\times a_{m_{q4}}^{\left( q\lambda \right) }\right] ^{\left( \kappa
_2\sigma _2\right) }||nl^{N^{\prime }}\;\alpha ^{\prime }Q^{\prime
}L^{\prime }S^{\prime }).
\end{array}
\end{equation}

Schematically we can express the matrix element in the second case, when the
operators of second quantization act upon the second shell, as follows:

\begin{equation}
\label{eq:ggbb}
\begin{array}[b]{c}
(n_1l_1^{N_1}n_2l_2^{N_2}\alpha _1L_1S_1Q_1M_{Q_1}\alpha
_2L_2S_2Q_2M_{Q_2}LSM_LM_S|G(2222) \\
|n_1l_1^{N_1}n_2l_2^{N_2}\alpha _1^{\prime }L_1^{\prime }S_1^{\prime
}Q_1^{\prime }M_{Q_1}\alpha _2^{\prime }L_2^{\prime }S_2^{\prime
}Q_2^{\prime }M_{Q_2}L^{\prime }S^{\prime }M_L^{\prime }M_S^{\prime })= \\
=
\displaystyle {\sum_{\kappa _{12},\sigma _{12},\kappa _{12}^{\prime }\sigma
_{12}^{\prime }}}\Theta \left( \kappa _{12},\sigma _{12},\kappa
_{12}^{\prime },\sigma _{12}^{\prime },n_2,\lambda _2\right) \times \\
\times (n_1l_1^{N_1}n_2l_2^{N_2}\alpha _1L_1S_1Q_1M_{Q_1}\alpha
_2L_2S_2Q_2M_{Q_2}LSM_LM_S| \\
A_{p-p}^{\left( kk\right) }\left( \kappa _{12},\sigma _{12},\kappa
_{12}^{\prime },\sigma _{12}^{\prime },n_2,\lambda _2\right) \\
|n_1l_1^{N_1}n_2l_2^{N_2}\alpha _1^{\prime }L_1^{\prime }S_1^{\prime
}Q_1^{\prime }M_{Q_1}\alpha _2^{\prime }L_2^{\prime }S_2^{\prime
}Q_2^{\prime }M_{Q_2}L^{\prime }S^{\prime }M_L^{\prime }M_S^{\prime }),
\end{array}
\end{equation}
and find its value by using Wigner-Eckart theorem, expressions (4.9), (2.28)
from Jucys and Savukynas 1973 as well as (\ref{eq:gj}) and (\ref{eq:gk}).

Differently from the first and the second cases, in the third ($n_i\lambda
_i=n_i^{\prime }\lambda _i^{\prime }=n_1l_1s,\;n_j\lambda _j=n_j^{\prime
}\lambda _j^{\prime }=n_2l_2s$) and the fourth ($n_i\lambda _i=n_i^{\prime
}\lambda _i^{\prime }=n_2l_2s,\;n_j\lambda _j=n_j^{\prime }\lambda
_j^{\prime }=n_1l_1s$) cases the first tensorial form (\ref{eq:gb}) is not
convenient for calculating the matrix elements. This is related to the fact
that the spin-angular part of matrix elements do not have shape of any
expression below:

\begin{equation}
\label{eq:ggga}
\begin{array}[b]{c}
(n_1l_1^{N_1}n_2l_2^{N_2}\alpha _1L_1S_1Q_1M_{Q_1}\alpha
_2L_2S_2Q_2M_{Q_2}LS|| \\
A^{\left( kk\right) }\left( \kappa _{12},\sigma _{12},\kappa _{12}^{\prime
},\sigma _{12}^{\prime },n_1,\lambda _1\right) \\
||n_1l_1^{N_1}n_2l_2^{N_2}\alpha _1^{\prime }L_1^{\prime }S_1^{\prime
}Q_1^{\prime }M_{Q_1}\alpha _2^{\prime }L_2^{\prime }S_2^{\prime
}Q_2^{\prime }M_{Q_2}L^{\prime }S^{\prime }),
\end{array}
\end{equation}

\begin{equation}
\label{eq:gggb}
\begin{array}[b]{c}
(n_1l_1^{N_1}n_2l_2^{N_2}\alpha _1L_1S_1Q_1M_{Q_1}\alpha
_2L_2S_2Q_2M_{Q_2}LS|| \\
A^{\left( kk\right) }\left( \kappa _{12},\sigma _{12},\kappa _{12}^{\prime
},\sigma _{12}^{\prime },n_2,\lambda _2\right) \\
||n_1l_1^{N_1}n_2l_2^{N_2}\alpha _1^{\prime }L_1^{\prime }S_1^{\prime
}Q_1^{\prime }M_{Q_1}\alpha _2^{\prime }L_2^{\prime }S_2^{\prime
}Q_2^{\prime }M_{Q_2}L^{\prime }S^{\prime }),
\end{array}
\end{equation}

\begin{equation}
\label{eq:gggc}
\begin{array}[b]{c}
(n_1l_1^{N_1}n_2l_2^{N_2}\alpha _1L_1S_1Q_1M_{Q_1}\alpha
_2L_2S_2Q_2M_{Q_2}LS|| \\
\left[ A^{\left( \kappa _{12}\sigma _{12}\right) }\left( n_1\lambda
_1\right) \times B^{\left( \kappa _{12}^{\prime }\sigma _{12}^{\prime
}\right) }\left( n_2\lambda _2\right) \right] ^{\left( kk\right) } \\
||n_1l_1^{N_1^{\prime }}n_2l_2^{N_2^{\prime }}\alpha _1^{\prime }L_1^{\prime
}S_1^{\prime }Q_1^{\prime }M_{Q_1}^{\prime }\alpha _2^{\prime }L_2^{\prime
}S_2^{\prime }Q_2^{\prime }M_{Q_2}^{\prime }L^{\prime }S^{\prime }),
\end{array}
\end{equation}

\begin{equation}
\label{eq:gggd}
\begin{array}[b]{c}
(n_1l_1^{N_1}n_2l_2^{N_2}\alpha _1L_1S_1Q_1M_{Q_1}\alpha
_2L_2S_2Q_2M_{Q_2}LS|| \\
\left[ A^{\left( \kappa _{12}\sigma _{12}\right) }\left( n_2\lambda
_2\right) \times B^{\left( \kappa _{12}^{\prime }\sigma _{12}^{\prime
}\right) }\left( n_1\lambda _1\right) \right] ^{\left( kk\right) } \\
||n_1l_1^{N_1^{\prime }}n_2l_2^{N_2^{\prime }}\alpha _1^{\prime }L_1^{\prime
}S_1^{\prime }Q_1^{\prime }M_{Q_1}^{\prime }\alpha _2^{\prime }L_2^{\prime
}S_2^{\prime }Q_2^{\prime }M_{Q_2}^{\prime }L^{\prime }S^{\prime }).
\end{array}
\end{equation}

Here $A^{\left( \kappa _{12}\sigma _{12}\right) }\left( n\lambda \right) $
and $B^{\left( \kappa _{12}^{\prime }\sigma _{12}^{\prime }\right) }\left(
n\lambda \right) $ represent any tensorial operator.

Only these shapes (\ref{eq:ggga}), (\ref{eq:gggb}), (\ref{eq:gggc}), (\ref
{eq:gggd}), in the case of two open shells, guarantee the effective use of
Racah algebra. That includes the determination of zero matrix elements from
triangular conditions (for example in (\ref{eq:gggc}) $\delta \left(
L_1,L_1^{\prime },\kappa _{12}\right) $, $\delta \left( S_1,S_1^{\prime
},\sigma _{12}\right) $, $\delta \left( L_2,L_2^{\prime },\kappa
_{12}^{\prime }\right) $, $\delta \left( S_2,S_2^{\prime },\sigma
_{12}^{\prime }\right) $) without explicit calculation, the use of tables of
standard quantities, and the use of quasispin (see Section 4) at last.

Meanwhile the second form (\ref{eq:gc}) allows one to exploit the Racah
algebra to its full extent, as the matrix elements for third case

\begin{equation}
\label{eq:ggc}
\begin{array}[b]{c}
(n_1l_1^{N_1}n_2l_2^{N_2}\alpha _1L_1S_1Q_1M_{Q_1}\alpha
_2L_2S_2Q_2M_{Q_2}LSM_LM_S|G_{II}(1212) \\
|n_1l_1^{N_1}n_2l_2^{N_2}\alpha _1^{\prime }L_1^{\prime }S_1^{\prime
}Q_1^{\prime }M_{Q_1}\alpha _2^{\prime }L_2^{\prime }S_2^{\prime
}Q_2^{\prime }M_{Q_2}L^{\prime }S^{\prime }M_L^{\prime }M_S^{\prime })= \\
=\frac 12
\displaystyle {\sum_p}\left( -1\right) ^{k-p}\left[ \kappa _1,\kappa
_2,\sigma _1,\sigma _2\right] ^{-1/2}\times \\ \times \left( n_1\lambda
_1n_2\lambda _2||g^{\left( \kappa _1\kappa _2k,\sigma _1\sigma _2k\right)
}||n_1\lambda _1n_2\lambda _2\right) \times \\
\times (n_1l_1^{N_1}n_2l_2^{N_2}\alpha _1L_1S_1Q_1M_{Q_1}\alpha
_2L_2S_2Q_2M_{Q_2}LSM_LM_S| \\
\left[ \left[ a^{\left( \lambda _1\right) }\times
\stackrel{\sim }{a}^{\left( \lambda _1\right) }\right] ^{\left( \kappa
_1\sigma _1\right) }\times \left[ a^{\left( \lambda _2\right) }\times
\stackrel{\sim }{a}^{\left( \lambda _2\right) }\right] ^{\left( \kappa
_2\sigma _2\right) }\right] _{p-p}^{\left( kk\right) } \\
|n_1l_1^{N_1}n_2l_2^{N_2}\alpha _1^{\prime }L_1^{\prime }S_1^{\prime
}Q_1^{\prime }M_{Q_1}\alpha _2^{\prime }L_2^{\prime }S_2^{\prime
}Q_2^{\prime }M_{Q_2}L^{\prime }S^{\prime }M_L^{\prime }M_S^{\prime }).
\end{array}
\end{equation}
and fourth case

\begin{equation}
\label{eq:ggcc}
\begin{array}[b]{c}
(n_1l_1^{N_1}n_2l_2^{N_2}\alpha _1L_1S_1Q_1M_{Q_1}\alpha
_2L_2S_2Q_2M_{Q_2}LSM_LM_S|G_{II}(2121) \\
|n_1l_1^{N_1}n_2l_2^{N_2}\alpha _1^{\prime }L_1^{\prime }S_1^{\prime
}Q_1^{\prime }M_{Q_1}\alpha _2^{\prime }L_2^{\prime }S_2^{\prime
}Q_2^{\prime }M_{Q_2}L^{\prime }S^{\prime }M_L^{\prime }M_S^{\prime })= \\
=\frac 12
\displaystyle {\sum_p}\left( -1\right) ^{k-p}\left[ \kappa _1,\kappa
_2,\sigma _1,\sigma _2\right] ^{-1/2}\times \\ \times \left( n_2\lambda
_2n_1\lambda _1||g^{\left( \kappa _1\kappa _2k,\sigma _1\sigma _2k\right)
}||n_2\lambda _2n_1\lambda _1\right) \times \\
\times (n_1l_1^{N_1}n_2l_2^{N_2}\alpha _1L_1S_1Q_1M_{Q_1}\alpha
_2L_2S_2Q_2M_{Q_2}LSM_LM_S| \\
\left[ \left[ a^{\left( \lambda _2\right) }\times
\stackrel{\sim }{a}^{\left( \lambda _2\right) }\right] ^{\left( \kappa
_1\sigma _1\right) }\times \left[ a^{\left( \lambda _1\right) }\times
\stackrel{\sim }{a}^{\left( \lambda _1\right) }\right] ^{\left( \kappa
_2\sigma _2\right) }\right] _{p-p}^{\left( kk\right) } \\
|n_1l_1^{N_1}n_2l_2^{N_2}\alpha _1^{\prime }L_1^{\prime }S_1^{\prime
}Q_1^{\prime }M_{Q_1}\alpha _2^{\prime }L_2^{\prime }S_2^{\prime
}Q_2^{\prime }M_{Q_2}L^{\prime }S^{\prime }M_L^{\prime }M_S^{\prime }).
\end{array}
\end{equation}
are schematically written down in a following as (\ref{eq:gggc}) and (\ref
{eq:gggd}), by using expression (4.3) from Jucys and Savukynas 1973,

\begin{equation}
\label{js-c}
\begin{array}{c}
(\alpha _1j_1\alpha _2j_2j||\left[ A_1^{\left( k_1\right) }\times
A_2^{\left( k_2\right) }\right] ^{\left( k\right) }||\alpha _1^{\prime
}j_1^{\prime }\alpha _2^{\prime }j_2^{\prime }j^{\prime })=\left[
j,j^{\prime },k\right] ^{1/2}\times \\
\times (\alpha _1j_1||A_1^{\left( k_1\right) }||\alpha _1^{\prime
}j_1^{\prime })(\alpha _2j_2||A_2^{\left( k_2\right) }||\alpha _2^{\prime
}j_2^{\prime })\left\{
\begin{array}{ccc}
j_1 & j_2 & j \\
j_1^{\prime } & j_2^{\prime } & j^{\prime } \\
k_1 & k_2 & k
\end{array}
\right\}
\end{array}
\end{equation}
and in the fourth case, after reversing the order of shells and altering the
coupling of their momenta for bra and ket functions we obtain:

\begin{equation}
\label{eq:ggd}
\begin{array}[b]{c}
(n_1l_1^{N_1}n_2l_2^{N_2}\alpha _1L_1S_1Q_1M_{Q_1}\alpha
_2L_2S_2Q_2M_{Q_2}LS|| \\
\left[ \left[ a^{\left( \lambda _1\right) }\times
\stackrel{\sim }{a}^{\left( \lambda _1\right) }\right] ^{\left( \kappa
_1\sigma _1\right) }\times \left[ a^{\left( \lambda _2\right) }\times
\stackrel{\sim }{a}^{\left( \lambda _2\right) }\right] ^{\left( \kappa
_2\sigma _2\right) }\right] ^{\left( kk\right) } \\
||n_1l_1^{N_1}n_2l_2^{N_2}\alpha _1^{\prime }L_1^{\prime }S_1^{\prime
}Q_1^{\prime }M_{Q_1}\alpha _2^{\prime }L_2^{\prime }S_2^{\prime
}Q_2^{\prime }M_{Q_2}L^{\prime }S^{\prime })= \\
=\left[ k\right] \left[ L,S,L^{\prime },S^{\prime }\right] ^{1/2}\left\{
\begin{array}{ccc}
L_1 & L_2 & L \\
L_1^{\prime } & L_2^{\prime } & L^{\prime } \\
\kappa _1 & \kappa _2 & k
\end{array}
\right\} \left\{
\begin{array}{ccc}
S_1 & S_2 & S \\
S_1^{\prime } & S_2^{\prime } & S^{\prime } \\
\sigma _1 & \sigma _2 & k
\end{array}
\right\} \times \\
\times (n_1l_1^{N_1}\;\alpha _1Q_1L_1S_1||\left[ a^{\left( \lambda _1\right)
}\times
\stackrel{\sim }{a}^{\left( \lambda _1\right) }\right] ^{\left( \kappa
_1\sigma _1\right) }||n_1l_1^{N_1}\;\alpha _1^{\prime }Q_1^{\prime
}L_1^{\prime }S_1^{\prime })\times \\ \times (n_2l_2^{N_2}\;\alpha
_2Q_2L_2S_2||\left[ a^{\left( \lambda _2\right) }\times \stackrel{\sim }{a}%
^{\left( \lambda _2\right) }\right] ^{\left( \kappa _2\sigma _2\right)
}||n_2l_2^{N_2}\;\alpha _2^{\prime }Q_2^{\prime }L_2^{\prime }S_2^{\prime
}),
\end{array}
\end{equation}

\begin{equation}
\label{eq:ggdd}
\begin{array}[b]{c}
(n_1l_1^{N_1}n_2l_2^{N_2}\alpha _1L_1S_1Q_1M_{Q_1}\alpha
_2L_2S_2Q_2M_{Q_2}LS|| \\
\left[ \left[ a^{\left( \lambda _2\right) }\times
\stackrel{\sim }{a}^{\left( \lambda _2\right) }\right] ^{\left( \kappa
_1\sigma _1\right) }\times \left[ a^{\left( \lambda _1\right) }\times
\stackrel{\sim }{a}^{\left( \lambda _1\right) }\right] ^{\left( \kappa
_2\sigma _2\right) }\right] ^{\left( kk\right) } \\
||n_1l_1^{N_1}n_2l_2^{N_2}\alpha _1^{\prime }L_1^{\prime }S_1^{\prime
}Q_1^{\prime }M_{Q_1}\alpha _2^{\prime }L_2^{\prime }S_2^{\prime
}Q_2^{\prime }M_{Q_2}L^{\prime }S^{\prime })= \\
=\left[ k\right] \left[ L,S,L^{\prime },S^{\prime }\right] ^{1/2}\left\{
\begin{array}{ccc}
L_2 & L_1 & L \\
L_2^{\prime } & L_1^{\prime } & L^{\prime } \\
\kappa _1 & \kappa _2 & k
\end{array}
\right\} \left\{
\begin{array}{ccc}
S_2 & S_1 & S \\
S_2^{\prime } & S_1^{\prime } & S^{\prime } \\
\sigma _1 & \sigma _2 & k
\end{array}
\right\} \times \\
\times (n_2l_2^{N_2}\;\alpha _2Q_2L_2S_2||\left[ a^{\left( \lambda _2\right)
}\times
\stackrel{\sim }{a}^{\left( \lambda _2\right) }\right] ^{\left( \kappa
_1\sigma _1\right) }||n_2l_2^{N_2}\;\alpha _2^{\prime }Q_2^{\prime
}L_2^{\prime }S_2^{\prime })\times \\ \times (n_1l_1^{N_1}\;\alpha
_1Q_1L_1S_1||\left[ a^{\left( \lambda _1\right) }\times \stackrel{\sim }{a}%
^{\left( \lambda _1\right) }\right] ^{\left( \kappa _2\sigma _2\right)
}||n_1l_1^{N_1}\;\alpha _1^{\prime }Q_1^{\prime }L_1^{\prime }S_1^{\prime
}),
\end{array}
\end{equation}

From this we conclude that in the third and fourth cases the usage of the
tensorial expressions of only two-particle operator (\ref{eq:gc}) allows us
to successfully exploit all the advantages of Racah algebra and quasispin
formalism in calculating the spin-angular parts of any two-particle operator
matrix element. This, to our mind, not only simplifies the calculations
considerably, by allowing to use the tables of irreducible tensors that are
independent of shell occupation numbers, but also allows one to establish
the zero matrix elements without performing explicit calculation.

Meanwhile the situation is different when the last two cases are considered,
or the matrix elements between more complex configurations are to be
established. This is related to the fact that using first (\ref{eq:gb}) or
second (\ref{eq:gc}) tensorial forms the spin-angular part of matrix
elements for those cases do not have shape of any expression (\ref{eq:ggga}%
), (\ref{eq:gggb}), (\ref{eq:gggc}) and (\ref{eq:gggd}).

In the next paper we shall present a methodology that allows one to use
efficiently the Racah algebra and quasispin formalism in a general case, too.

\section{\bf Conclusion}

Preliminary usage of the generalized graphical method, irreducible tensorial
form of the second quantization operators as well as of quasispin technique,
while calculating the spin-angular parts of matrix elements of the energy
operator, has demonstrated high efficiency to obtain in a uniform way the
general expressions for the operators of physical quantities as well as for
their matrix elements, covering the both cases of diagonal and non-diagonal
ones with respect to quantum numbers of electronic configurations. Therefore
it is fairly promising to formulate this methodology in a complete and
consistent way for an arbitrary number of electronic shells with its
successive implementation in the universal computer codes.

\section*{\bf Acknowledgements}

The authors are grateful to Professor Charlotte Froese Fischer for
encouraging and valuable remarks. This work is part of co-operative research
project funded by National Science Foundation under grant No. PHY-9501830
and by EURONET PECAM associated contract ERBCIPDCT 940025.

\clearpage

\section*{\bf References}


\begin{quote}
Cowan R D 1981 {\it The Theory of Atomic Structure and Spectra} (Berkeley,
CA: University of California Press)

Froese Fischer C 1977 {\it The Hartree-Fock Method for Atoms} (New York:
Wiley)

Gaigalas G A 1985 {\it Spectroscopy of autoionized states of atoms and ions}
(Moscow: Scientific council of spectroscopy) 43 (in Russian)

Gaigalas G A, Kaniauskas J M and Rudzikas Z B 1985 {\it Liet. Fiz. Rink.
(Sov. Phys. Collection)} {\bf 25} 3

Gaigalas G A and Merkelis G V 1987 {\it Acta Phys. Hungarica} {\bf 61}, 111

Jucys A P and Bandzaitis A\ A 1977 {\it Theory of Angular Momentum in
Quantum Mechanics} (Mokslas: Vilnius) (in Russian).

Jucys A P and Savukynas A J 1973 {\it Mathematical Foundations of the Atomic
Theory} (Vilnius: Mokslas) (in Russian)

Judd B R 1967 {\it Second Quantization and Atomic Spectroscopy} (Baltimore:
John Hopkins Press)

Lindgren I and Morrison M 1982 {\it Atomic Many-Body Theory (Springer Series
in Chemical Physics, vol. 13)} (Berlin: Springer-Verlag) (2nd edition)

Merkelis G V, Gaigalas G A, Kaniauskas J M and Rudzikas Z B 1986 {\it Izv.
AN\ SSSR Ser. Fiz.} {\bf 50} 1403

Merkelis G V, Gaigalas G A and Rudzikas Z B 1986 {\it Liet.fiz. rink. (Sov.
Phys. Collection)} {\bf 25},{\bf \ }14

Nikitin A A and Rudzikas Z B 1983 {\it Foundations of the Theory of the
Spectra of Atoms and Ions} (Moscow: Nauka) (in Russian)

Rudzikas Z B 1991 {\it Comments At. Mol. Phys.} {\bf 26} 269

Rudzikas Z B 1996 {\it Theoretical Atomic Spectroscopy (Many-Electron Atom)}
(Cambridge: Cambridge University Press) (in press)

Rudzikas Z B and Kaniauskas J M 1984 {\it Quasispin and Isospin in the
Theory of Atom} (Vilnius: Mokslas) (in Russian)


\end{quote}

\end{document}